\documentclass[final]{IEEEtran}

\IEEEoverridecommandlockouts
\usepackage{color}
\usepackage{amsmath}
\usepackage{amsfonts}
\usepackage{amssymb}
\usepackage{amscd}
\usepackage[]{graphicx}
\usepackage{newlfont}
\usepackage{cite}
\usepackage{subfigure}
\usepackage{soul}
\newcommand{\prob}{\mathcal{P}}

\newcommand{\lb}{\left(}
\newcommand{\rb}{\right)}
\newcommand{\lsqb}{\left[}
\newcommand{\rsqb}{\right]}
\newcommand{\lcb}{\left\{}
\newcommand{\rcb}{\right\}}

\newcommand{\honesq}{|h_1|^2}
\newcommand{\htwosq}{|h_2|^2}
\newcommand{\honetwosq}{|h_{12}|^2}
\newcommand{\hself}{g^2}
\newcommand{\Ronetermone}{\exp\lb -\gamma_1 d_1^\alpha(1 + g^2 P_J)\lb \frac{1}{P_1 - \gamma_1 P_2} - \frac{1}{P_1}\rb\rb}
\newcommand{\Ronetermthree}{\lsqb 1 + \gamma_1 \frac{P_J}{P_1} \lb \frac{d_2}{d_3} \rb^\alpha\rsqb}
\newcommand{\Ronetermtwo}{1 - \exp\lb -\frac{\gamma_1 d_2^\alpha}{P_1}\rb\Ronetermthree^{-1}}
\newcommand{\Ronetermfive}{\lsqb 1 + \gamma_1 \frac{P_J}{P_1 - \gamma_1 P_2}\lb \frac{d_2}{d_3}\rb^\alpha\rsqb}
\newcommand{\Ronetermfour}{1 - \exp\lb - \frac{\gamma_1 d_2^\alpha}{P_1 - \gamma_1 P_2}\rb\Ronetermfive^{-1}}
\newcommand{\Ronetermsix}{\exp\lb - \frac{\gamma_2 d_2^\alpha}{P_2 - \gamma_2 P_1}\rb \lsqb 1 + \gamma_2 \frac{P_J}{P_2 - \gamma_2 P_1} \lb \frac{d_2}{d_3}\rb^\alpha\rsqb^{-1}}
\newcommand{\Ronetermseven}{1 + \gamma_2 \frac{P_J}{P_2 - \gamma_2 P_1} \lb \frac{d_2}{d_3}\rb^\alpha}

\newcommand{\Rtwotermone}{1 + \gamma_2 \frac{P_J}{P_2 - \gamma_2 P_1} \lb \frac{d_2}{d_3}\rb^{\alpha}}
\newcommand{\Rtwotermtwo}{1 + \gamma_2 \frac{P_J}{P_2} \lb \frac{d_2}{d_3}\rb^\alpha}
\newcommand{\Rtwotermthree}{1 + \gamma_1 \frac{P_J}{P_1 - \gamma_1 P_2}\lb \frac{d_2}{d_3}\rb^\alpha}
\newcommand{\Rtwotermfour}{1 - \exp\lb - \frac{\gamma_1 d_2^\alpha}{P_2 - \gamma_2 P_1}\rb\lsqb \Rtwotermthree \rsqb^{-1}}

\ifCLASSINFOpdf
\else
\fi

\begin{document}
\title{Secure Communications for the Two-user Broadcast Channel with Random Traffic}
\author{
	\authorblockN{Parthajit~Mohapatra, Nikolaos~Pappas, \textit{Member, IEEE},\\ Jemin Lee, \textit{Member, IEEE}, ~Tony Q. S. Quek, \textit{Senior Member, IEEE}, \\ and Vangelis~Angelakis, \textit{Senior Member, IEEE}}
	
\thanks{Parthajit Mohapatra is with the G. S. Sanyal School of Telecommunications, Indian Institute of Technology, Kharagpur, India (e-mail: parthajit@gssst.iitkgp.ernet.in). 

Nikolaos~Pappas and  Vangelis~Angelakis are with the Department of Science and Technology, Link\"{o}ping University, Norrk\"{o}ping SE-60174, Sweden (e-mail: \{nikolaos.pappas, vangelis.angelakis\}@liu.se).

Jemin Lee is with Department of Information and Communication Engineering, Daegu Gyeongbuk Institute of Science and Technology (DGIST), Korea (e-mail: jmnlee@dgist.ac.kr). 

Tony Q. S. Quek is with the Information Systems Technology and Design Pillar, Singapore University of Technology and Design, Singapore (e-mail: tonyquek@sutd.edu.sg).

This work was presented in part in IEEE ICC 2016 \cite{Parthajit-iccBC-2016}.

The corresponding author is J. Lee. 
}}
\maketitle
\begin{abstract}
In this work, we study the stability region of the two-user broadcast channel (BC) with bursty data arrivals and security constraints. We consider the scenario, where one of the receivers has a secrecy constraint and its packets need to be kept secret from the other receiver. This is achieved by employing full-duplexing at the receiver with the secrecy constraint, so that it transmits a jamming signal to impede the reception of the other receiver. In this context, the stability region of the two-user BC is characterized for the general decoding case. Then, assuming two different decoding schemes the respective stability regions are derived. The effect of self-interference due to the full-duplex operation on the stability region is also investigated. The stability region of the BC with a secrecy constraint, where the receivers do not have full duplex capability can be obtained as a special case of the results derived in this paper. In addition, the paper considers the problem of maximizing the saturated throughput of the queue, whose packets does not require to be kept secret under minimum service guarantees for the other queue. The results provide new insights on the effect of the secrecy constraint on the stability region of the BC. In particular, it is shown that the stability region with secrecy constraint is sensitive to the coefficient of self-interference cancelation under certain cases. 
\end{abstract}

\section{Introduction}
In the last decade, physical layer secrecy has emerged as a promising approach for security in wireless communications. The physical layer techniques inspired from the information theoretic results often require infinitely backlogged users, i.e., users that always have data to transmit. However, such an assumption does not capture the bursty nature of the sources. To account for the bursty nature of the sources, the notion of stability region \cite{rao-TIT-1988} has been introduced. In many communication scenarios, it is required to serve multiple users simultaneously and the data arrival at the transmitter is bursty in nature. In addition, users may have different levels of security requirements. To examine jointly these two important aspects of communication, we consider a two-user broadcast channel (BC)~\cite{cover-TIT-1975} with bursty packet arrivals, where the packets of one of the queues need to be kept secret from the unintended receiver. We further assume that the receiver with the secrecy constraint has full-duplex capability. Thus, the full-duplex receiver can send a jamming signal to impede the decoding of its packets by the other receiver. The effect of the secrecy constraint on the stability region is not yet well understood. Furthermore, the stability of the queues can be affected due to the secrecy constraint and the self-interference caused because of the simultaneous transmission and reception at the receiver. Hence, the analysis of this model can provide useful insights on the system performance as well as help to understand the effect of bursty nature of sources in a broadcast channel with secrecy constraint.
 
\subsection{Related Work}
The capacity of the BC has been analyzed extensively with and without secrecy constraints in the existing literature \cite{cover-TIT-1975, marton-TIT-1979, jafarian-isit-2011, csiszar-TIT-1978, ekrem-TIT-2011}. However, the capacity region of the general BC is still unknown even without a secrecy constraint at the receiver. The achievable rate region derived in \cite{csiszar-TIT-1978} is the best known achievable rate region for a general discrete memoryless BC. The work in \cite{jafarian-isit-2011} provided a partial characterization of the capacity region of the two-user Gaussian fading broadcast channel. 

Exploiting randomness in a physical channel  to ensure secrecy was first considered the case of the wiretap channel, where the legitimate transmitter needs to send a message to the legitimate receiver securely in the presence of an eavesdropper \cite{wyner-bell-1975}. It was shown that secure communication is possible between the legitimate users without using a secret key between the legitimate nodes. The result of the wiretap channel was generalized in \cite{csiszar-TIT-1978} for the BC, where one of the messages needs to be kept secret from the unintended receiver. The effect of user cooperation on the secrecy capacity of the BC has been investigated in \cite{ekrem-TIT-2011}. The problem of secure broadcasting over fading channel has been considered in \cite{khisti-TIT-2008}. The results related to other multiuser scenarios with secrecy constraints can be found in \cite{liu-TIT-2008, lgamal-TIT-2011, ekrem2}. 

In wireless networks, the arrivals of data at the transmitter are random, and the vast majority of the works in information theory have the assumption of backlogged users \cite{wyner-bell-1975, csiszar-TIT-1978, ekrem-TIT-2011, barros-isit-2006, liu-TIT-2008, lgamal-TIT-2011, ekrem2}. When the traffic is bursty in nature, the \emph{stability region} or \emph{stable throughput region} becomes an appropriate measure of rates in packets/slot in wireless networks \cite{anthony-TIT-1998}. The stability region is defined as the set of all arrival rates for which all queues inside the network are stable, i.e., the length for each queues is finite~\cite{rao-TIT-1988}. The work in \cite{fayolle-acm-1977} provided a theoretical treatment of some basic problems related to stability of the broadcast channel. The stability region has been studied for several other communication models, such as the two-user interference channel \cite{pappas-itw-2013-IC}. The main difficulty in the derivation of the stability region is the interaction among the queues, so stochastic dominance \cite{rao-TIT-1988} was used to overcome this difficulty. However, the characterization of the stability region above three users remains a challenging problem. A detailed treatment of stability region issues and derivations can can be found in \cite{Sastry-NOW}.

The stability region has been analyzed for multiple users scenarios with secrecy constraints \cite{sarikaya-allerton-2009, Poor-ITA2008, Poor-TIFS2011}. In \cite{sarikaya-allerton-2009}, stability conditions are obtained for a slotted ALOHA network with secrecy constraints. In this case, all users who are not transmitting are the eavesdroppers. A wireless broadcast network model with secrecy constraints is investigated in  \cite{Poor-ITA2008, Poor-TIFS2011}, where a source node broadcasts confidential message to user nodes, where each message is required to be decoded by the intended user and to be kept secret from all other users. In \cite{Poor-ITA2008, Poor-TIFS2011}, secrecy, reliability, and stability are jointly considered for network utility maximization. 

\subsection{Contributions}
In this paper, the stable throughput region of the BC with a secrecy constraint satisfied by means of a full duplex receiver and under bursty traffic is analyzed. The stable throughput of this scenario has not been considered in the existing literature. The transmitter has two queues and the packet sent from the first and the second queues are for receiver~$1$ and receiver $2$, respectively. The packets for receiver $1$ need to be kept secret from the receiver $2$. 
In practice, such scenario can arise in a cellular network, where users have different subscription options, which produces different levels of data secrecy for users. Another practical scenario can appear in  IoT networks where there are sensors collecting data from the environment. The collected data can be categorized to non-confidential  ones (e.g., outdoors temperature) or confidential ones (e.g., a baby monitor feed). The data collecting node can store the confidential data and the non-confidential to different queues. The data are transmitted through a wireless channel to two different receivers, the first one requests the confidential data while the second one requests non-confidential ones.
For better data confidentially, the receiver 1 has the full duplex capability to generate jamming signals to receiver 2 while receiving packets from the transmitter.
In this case, the derivation of the stability region is challenging since there is a correlation between the queues, and the reliability and security criteria need to be taken into account jointly. 
Furthermore, differently to the works in \cite{Poor-ITA2008, Poor-TIFS2011}, which assume all the nodes know the channel state information,
we assume that channel state information is not available at the source node who only knows the statistics of channels.
The main contributions of this work can be summarized as follows.
\begin{enumerate}
\item The stability region of the two-user BC with the secrecy constraint is obtained by considering different decoding capabilities at the receivers. 
Specifically, we consider the case of decoding by treating interference as noise (i.e., limited decoding capability case) and performing successive decoding (i.e., high decoding capability case) for each of the receivers. The stability regions are derived using the stochastic dominance technique after determining the success probabilities with and without secrecy constraints for the different decoding schemes. The derivation of the success probability is nontrivial as it is required to satisfy also the secrecy criteria at the receivers. 

\item The self-interference cancelation capability at receiver 1 is also taken into account in the analysis of the stability region. 
Although receiver $1$ transmits a jamming signal to receiver $2$ to enhance the data confidentiality, 
it can also degrade its own throughput due to the self-interference. We explore how the self-interference cancelation capability affects the stability region. 
Furthermore, the stability region of the BC without FD at receiver 1 can be obtained as a special case of the results derived in this paper 
by setting the jamming power to zero.
\item The results derived in this paper are also useful to analyze other communication scenarios such as queues with different congestion levels. As one of such problems, where the jamming power is also optimized to maximize the saturated throughput of the second receiver while guaranteeing a required service rate for the first user.
\end{enumerate}
  
In comparison to our preliminary work presented in \cite{Parthajit-iccBC-2016}, the following additional results have been derived in this paper. In \cite{Parthajit-iccBC-2016}, it was assumed that receiver $D_2$ cannot perform successive decoding. Here, the stability region is also derived when receiver~$D_2$ can perform successive decoding and receiver~$D_1$ uses treats interference as noise while decoding its intended packet. The derivation also takes account of the fact that if receiver~$2$ fails to decode the packet of first user in the successive decoding, then it attempts to decode the packet of the other user by treating interference as noise. Using the results derived under different decoding assumptions, closure of the stability regions are obtained for two cases: (a) fixed powers at the transmitter, and (b) the jammer (i.e., $D_2$). Beside these, the problem of optimizing jamming power for maximizing the saturated throughput of second receiver while
guaranteeing a required service rate for the first user is also explored. 
\section{System Model}
We consider a two-user broadcast channel (BC) as shown in Fig.~\ref{eq:system-model1}, where a single transmitter has two different queues. The $i^{\text{th}}$ ($i =1, 2$) queue contains the packets intended to receiver~$i$ denoted also by $D_i, i = 1, 2$. There is no security constraint for the packets sent from the second queue, but the packets sent from the first queue are required to be kept secret from the second receiver, i.e., $D_2$. To enhance the communication secrecy, $D_1$ is assumed to have full-duplex capability, so it can receive the packets from the transmitter and send a jamming signal to the other receiver ($D_2$), simultaneously. The jamming signal sent by $D_1$ can cause additional interference at $D_2$ and may help to increase the secrecy throughput of $D_1$. The system model considered in this paper can capture a scenario, where one of the users requests for confidential data and another user requests for data without security requirements from the same transmitter.
\begin{figure}
	\centering
	\includegraphics[width=2.2in, ]{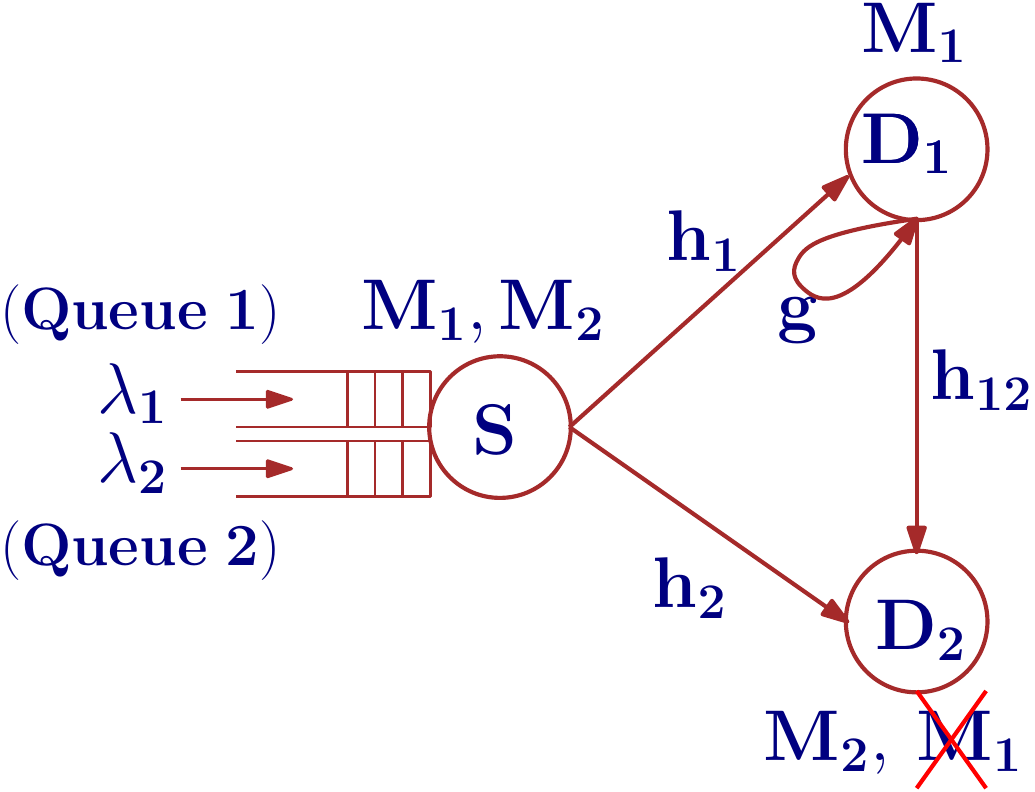}\\
	\caption{The two-user broadcast channel with the security constraint: receiver~$1$ has full-duplex capability.}\label{fig:system_model}
\end{figure}
The signal $y_i^{(t)}$ $(i \in \{1, 2\})$ at receiver~$i$ at time-slot $t$ is modeled as 
\begin{align}
& y_1^{(t)} = h_1^{(t)} x^{(t)} + g x_{\text{J}}^{(t)} + z_1^{(t)}, \nonumber \\
& y_2^{(t)} = h_2^{(t)} x^{(t)} + h_{12}^{(t)} x_{\text{J}}^{(t)} + z_2^{(t)}, \label{eq:system-model1}
\end{align}
where $z_i^{(t)} \sim \mathcal{CN}(0,1)$, and $h_i^{(t)}$ and $h_{ij}^{(t)}$ denote Rayleigh block fading channel between the transmitter and the $i$-th receiver and that between the $i$-th transmitter and the $j$-th receiver, respectively. When both the queues are non-empty, the transmitter sends $x^{(t)} = x_1^{(t)} + x_2^{(t)}$, where $x_i^{(t)}$ is the packet for receiver~$i$. When only the $i^{\text{th}}$ queue has packets to send, then the transmitter sends $x^{(t)} = x_i^{(t)}$ $(i \in \{1, 2\})$.

Here, it is assumed that all channels are independent. In \eqref{eq:system-model1}, $g$ is the residual self-interference, which is modeled by a scalar $g \in [0, 1]$ as for example in \cite{Weeraddana-wiopt-2010, Weeraddana-itw-2010, jlee-twcom-2015, pappas_twc15, nikos-access2017}. This captures the accuracy of the self-interference cancelation technique. When $g=1$, it indicates that no self-interference cancelation technique has been used and $g=0$ indicates that perfect self-interference cancelation has been achieved. Studying in detail the physical layer implementation or considering specific self-interference cancelation mechanisms is beyond the scope of this work.

The event $\mathcal{D}_{i/\mathcal{T}}$ denotes that the receiver~$i$ is able to decode the packet sent from the $i^{\text{th}}$ queue of the transmitter given a set of non-empty queues $\mathcal{T}$. The event $\mathcal{D}_{i/i, \mathcal{T}}^{s}$ denotes that the user $i$ is able to decode its intended packet and the other users $j \in \mathcal{T}~\backslash~\{i\}$  are not able to decode the packet sent from the $i^{\text{th}}$ queue. We use $s$ as superscript in $\mathcal{D}_{i/i, \mathcal{T}}^{s}$ as it maps the event that a packet for user~$i$ is confidential to other users. 

The packet arrival process at the $i$-th queue is assumed to be independent and stationary with mean rate $\lambda_i$ in packets per slot. Both the queues are assumed to have infinite capacity to store incoming packets (i.e., no loss system) and $Q_i$ denotes the length of the $i^{\text{th}}$ queue. If one of the queues is non-empty, then the transmitter sends a packet in a time slot. It is also assumed that acknowledgments sent by the receivers are instantaneous and error free. The average service rate for the first queue is
\begin{align}
\mu_1 \!= \!\prob \lb Q_2 > 0\rb \prob\lb  \mathcal{D}_{1/1,2}^s \rb  \!+ \! \prob \lb Q_2 = 0\rb \prob \lb \mathcal{D}_{1/1}^s\rb. \label{eq:system-model2}
\end{align}
Similarly, the average service rate for the second queue is
\begin{align}
\mu_2 = \prob \lb Q_1 > 0\rb \prob\lb  \mathcal{D}_{2/1,2} \rb + \prob \lb Q_1 = 0\rb 
\prob \lb \mathcal{D}_{2/2}\rb. \label{eq:system-model3}
\end{align}
Here, $\prob\lb\mathcal{A}\rb$ is the probability of having the event $\mathcal{A}$.
Hence, $\prob\lb\mathcal{D}_{i/\mathcal{T}}\rb$ and $\prob\lb\mathcal{D}_{i/i, \mathcal{T}}^{s}\rb$ denote
the probabilities that the receiver, $D_i$, decodes its intended packets successfully 
in the absence/presence of the secrecy constraint, respectively. 

\subsection{The stability region of the general case}
The following definition of queue stability is used \cite{Wojciech-1994-JSTOR}:\\
\textbf{Definition 1.} Let $Q_i^{(t)}$ denotes the length of the queue~$i$ at the beginning of the time-slot $t$. The queue is said to be stable if $\lim_{t \rightarrow \infty} \prob\lb Q_i^{(t)} < x\rb = F(x)$ and $\lim_{x \rightarrow \infty} F(x)= 1$. The queue is said to be sub-stable if 
$\lim_{x \rightarrow \infty} \lim_{t \rightarrow \infty} \inf \prob\lb Q_i^{(t)} < x \rb = 1$. If a queue is stable, then it is also sub-stable. A queue is said to be unstable if the queue is not sub-stable. 

From the definition of queue stability, the \emph{stability region} is defined as the set of all possible arrival rates at the queues for which the queues remain stable. The average service rates of the first and second queues are given by (\ref{eq:system-model2}) and (\ref{eq:system-model3}), respectively. Note that the average service rate of each queue depends on the queue length of the other queue, so it is non-trivial to determine the stability region. To overcome this difficulty, the stochastic dominance technique \cite{rao-TIT-1988} can be applied which is based on the construction of hypothetical dominant systems. The stability region $\mathcal{R}_{1}$ obtained from the first dominant system, i.e., when the first queue transmits dummy packets is as follows:

\begin{align}
\mathcal{R}_1 & = \lcb (\lambda_1, \lambda_2):\! \frac{\lambda_1}{\prob(\mathcal{D}_{1/1}^s)} \!+\! \frac{\prob(\mathcal{D}_{1/1}^s) -
	\prob(\mathcal{D}_{1/1,2}^s)}{\prob(\mathcal{D}_{1/1}^s)\: \prob(\mathcal{D}_{2/1,2})}\lambda_2 < 1, \right. \nonumber \\
& \qquad  \qquad \quad \lambda_2 < \prob(\mathcal{D}_{2/1,2}) \Bigg\}. \label{eq:stability-region1}
\end{align}
Similarly, the stability region $\mathcal{R}_{2}$ obtained from the second dominant system, 
i.e., when the second queue transmits dummy packets is as follows: 
\begin{align}
\mathcal{R}_2& = \lcb (\lambda_1, \lambda_2): \frac{\lambda_2}{\prob(\mathcal{D}_{2/2})} \!+\! \frac{\prob(\mathcal{D}_{2/2}) -
	\prob(\mathcal{D}_{2/1,2})}{\prob(\mathcal{D}_{2/2})\: \prob(\mathcal{D}_{1/1,2}^s)}\lambda_1 < 1, \right. \nonumber \\
& \qquad \qquad \quad  \lambda_1 < 
\prob(\mathcal{D}_{1/1,2}^s)\bigg\} \label{eq:stability-region2}
\end{align}

The stability region of the system is obtained by taking union of $\mathcal{R}_1$ and $\mathcal{R}_2$, i.e., $\mathcal{R} = \mathcal{R}_1 \cup \mathcal{R}_2$, where $\mathcal{R}_1$ and $\mathcal{R}_2$ corresponds to the stability regions of dominant systems in (\ref{eq:stability-region1}) 
and (\ref{eq:stability-region2}), respectively. The stability region above is expressed as a function of the success probabilities without the secrecy constraint i.e.,  $\prob\lb \mathcal{D}_{i/\mathcal{T}}\rb$, and with the secrecy constraint i.e., $\prob\lb \mathcal{D}_{i/i, \mathcal{T}}^{s}\rb$. The stability region determined using (\ref{eq:stability-region1}) and (\ref{eq:stability-region2}) is general, and no specific decoding needs to be assumed at the receivers. Note that the queue stability conditions are necessary and sufficient for the stability of the original system as obtained in \cite{pappas-arxiv-2015}, and the proof is based on \cite{rao-TIT-1988}.

\section{Stability Region Analysis with Different Decoding Schemes at the Receivers}\label{sec:stability-region-diff-decode}
In this section, the success probabilities with and without secrecy constraints are obtained with different decoding schemes at the receivers to determine the stability region. In a multiuser environment, users can have different decoding capabilities due to different hardware, and this in turn can affect the overall system performance. To study the impact of different decoding capabilities at the receiver, secrecy constraint, and self-interference on the stability region, the following cases are considered.

\begin{enumerate}	
\item Both the users have limited decoding capability and cannot perform successive decoding. In this case, both the users decode their intended packets with treating interference as noise, i.e., they treat other user's packet as noise while decoding its own packet.\footnote{From practical perspective, treating interference as noise is an attractive choice for decoding in many real world applications due to its low complexity and robustness to channel uncertainty.}

\item The receiver with full-duplex capability, i.e., $D_1$, can perform successive decoding while the other receiver $D_2$ treats interference as noise. In successive decoding, the receiver first tries to decode the packet of the unintended user, and then decodes its own packet after canceling the effect of the other user's packet. 

\item $D_1$ cannot perform successive decoding and decodes its intended packet by treating interference as noise while the second receiver $D_2$ can perform successive decoding. The second and third cases are of interest for the scenarios, where one of the receivers has limited decoding capability and the other receiver can perform successive decoding.\footnote{This is also relevant when one of the receivers is far from the transmitter compared to the other receiver and is in general true for the broadcast channel.}
\end{enumerate}

For the third case, it is assumed that if receiver~$2$ fails to decode the packet of first user in the successive decoding, then it attempts to decode the packet of the other user by treating interference as noise. Recall that the packet intended for receiver~$1$ needs to be kept secret from receiver~$2$ and this additional capability make receiver~$2$ more capable compared to receiver~$1$. In the following section, using the expressions for the success probabilities for different decoding schemes in (\ref{eq:stability-region1}) and (\ref{eq:stability-region2}), the stability region is obtained for the cases discussed above.

\subsection{Both the Receivers Treat Interference as Noise}\label{sec:decoding-cons-rx2}
In this case, it is assumed that both the receivers have limited decoding capabilities and cannot perform successive decoding. Hence, receiver~$1$ has to treat the received packet intended for receiver~$2$ and the residual jamming signal after self-interference cancelation as noise to decode its intended packet. Similarly, receiver~$2$ treats both the packets intended for receiver~$1$ and the jamming signal sent by receiver~$1$ as noise while decoding its intended packet. In the following, the success probability for both receivers are obtained with and without secrecy constraint for different status of the queues. We denote the threshold for decoding a packet at the receiver sent from the $i^{\text{th}}$ queue as $\gamma_i$ ($i=1,2$), the distance between the transmitter and the receiver~$i$ as $d_i$ $(i=1,2)$, the distance between the receivers as $d_3$, and the path loss exponent as $\alpha$.

\subsubsection{When $Q_1 = 0$ and $Q_2 \neq 0$} In this case, the first queue is empty, so only the second queue at the transmitter sends a packet for the receiver~$2$. Since there is no secrecy constraint for the packets intended to receiver~$2$ at queue~$2$, the receiver~$2$ can decode its intended packet if the following event is true
\begin{align}
 \mathcal{D}_{2/2} = \lcb P_2 \htwosq d_2^{-\alpha} \geq \gamma_2\rcb. \label{eq:fdasnoise1} 
\end{align}
Note that receiver~$1$ does not send a jamming signal in this case since receiver~$1$ can know in advance that it is not going to receive packets from the first queue through exchange of control signals with transmitter.  The success probability is then given by
\begin{align}
 \prob(\mathcal{D}_{2/2}) & = 1 - \prob\lcb P_2 \htwosq d_2^{-\alpha} < 
\gamma_2\rcb \nonumber \\
%New eqn
& = 1 - \int_{0}^{\infty} F_{\htwosq}\lb \frac{\gamma_2 }{P_2 d_2^{-\alpha}}\rb 
f_{\honetwosq}(x) dx \nonumber \\
%New eqn
& = \exp\lb -\frac{\gamma_2 d_2^\alpha}{P_2}\rb. \label{eq:fdasnoise2}
\end{align}
where $F_X(x)$ is the CDF of the random variable $X$ and $f_{\htwosq}(x) = \exp(-x)$.  
%New subsection
\subsubsection{When $Q_1 \neq 0$ and $Q_2 = 0$}
In this case, queue~$2$ is empty, so receiver~$2$ does not have a packet to decode from queue~$2$. However, receiver $2$ tries to decode the packet intended to receiver~$1$. Hence, there is an additional secrecy constraint compared to the previous case. The packet of queue~$1$ sent by transmitter should be decoded at receiver~$1$ while it should be kept confidential from receiver~$2$, and this is presented by the following event
\begin{align}
 \mathcal{D}_{1/1}^s = \lcb\frac{P_1 \honesq d_1^{-\alpha}}{1 + P_J \hself } \geq \gamma_1, 
  \frac{P_1 \htwosq d_2^{-\alpha}}{1 + P_J \honetwosq d_3^{-\alpha}} \leq 
  \gamma_1\rcb. \label{eq:fdasnoise3}
\end{align}

Note that in \cite{csiszar-TIT-1978}, the secrecy of the message is based on information theoretic approach. However, in this work we consider finite packet length, thus, the secrecy criteria is defined in terms of SINR as in \cite{Sarma-Wiopt2013, Bashar-Globecom2009, Bashar-MILCOM2009, Capar-INFOCOM2012, Jemin-JSAC2013}.

Since $h_1$, $h_2$ and $h_{12}$ are independent of each other, the success probability for receiver~$1$ is given by
\begin{align}
 & \prob(\mathcal{D}_{1/1}^s) \nonumber \\
 & = \prob \!\lcb \!\frac{P_1 \honesq d_1^{-\alpha}}{1 + P_J \hself} \geq \gamma_1 \!\rcb 
\!\!\prob\! \lcb \!\frac{P_1 \htwosq d_2^{-\alpha}}{1 + P_J \honetwosq d_3^{-\alpha}} \leq 
  \gamma_1 \!\rcb \nonumber \\
  %New eqn
  & \!= \!\exp\lb \!- \!\frac{\gamma_1 d_1^\alpha}{P_1} (1 \!+\! P_J g^2)\rb  \lcb 1 \!-\!  \exp\lb - \frac{\gamma_1d_2^\alpha}{P_1}\rb \right. \nonumber \\
 &  \left. \qquad \qquad  \times \lsqb 1 + 
\gamma_1 \frac{P_J}{P_1} \lb \frac{d_2}{d_3}\rb^\alpha\rsqb^{-1}\rcb. \label{eq:fdasnoise5}
\end{align}

\subsubsection{When $Q_1 \neq 0$ and $Q_2 \neq 0$}
In this case, both the queues have packets to transmit. The packet sent from the first queue should be decoded at receiver~$1$ and at the same time be confidential to receiver~$2$, which is presented by the following event
\begin{align}
  \mathcal{D}_{1/1,2}^s & = \lcb \frac{P_1 \honesq d_1^{-\alpha}}{1 + P_2 \honesq d_1^{-\alpha} + 
  P_J \hself } \geq \gamma_1, \right. \nonumber \\
& \left.  \qquad \frac{P_1 \htwosq d_2^{-\alpha}}{1 + P_2 \htwosq d_2^{-\alpha} + 
  P_J \honetwosq d_3^{-\alpha}} < \gamma_1\rcb. \label{eq:fdasnoise5a}
\end{align}
The above event is feasible, if the following condition is satisfied
\begin{align}
\frac{\honesq d_1^{-\alpha}}{1 + P_2 \honesq d_1^{-\alpha} + P_J g^2 } > \frac{\htwosq d_2^{-\alpha}}{1 + P_2 \htwosq d_2^{-\alpha} + P_J \honetwosq d_3^{-\alpha} 
}. \label{eq:fdasnoise5c}
\end{align}
The success probability in this case becomes
\begin{align}
&  \prob(\mathcal{D}_{1/1,2}^s) \nonumber \\
& = \prob\lcb \frac{P_1 \honesq d_1^{-\alpha}}{1 + P_2 \honesq d_1^{-\alpha} + 
  P_J \hself } \geq \gamma_1,  \right. \nonumber \\
& \qquad \left. \frac{P_1 \htwosq d_2^{-\alpha}}{1 + P_2 \htwosq d_2^{-\alpha} + 
  P_J \honetwosq d_3^{-\alpha}} < \gamma_1\rcb \nonumber \\
  %New eqn
  & = \prob \lcb \frac{(P_1 - \gamma_1 
  P_2)\honesq d_1^{-\alpha}}{1 + P_J \hself } \geq \gamma_1\rcb \times \nonumber \\
& \qquad \quad   \prob\ \lcb \frac{(P_1 - \gamma_1 
  P_2)\htwosq d_2^{-\alpha}}{1 + P_J \honetwosq d_3^{-\alpha}} < \gamma_1\rcb \nonumber \\
  %New eqn
  & = \exp\lb - \frac{\gamma_1 d_1^\alpha}{P_1 - \gamma_1 P_2} (1 + P_J g^2)\rb \lcb 1 -  
  \right. \nonumber \\
  & \quad  \left. \exp\lb - \frac{\gamma_1 d_2^\alpha}{P_1 - \gamma_1 P_2}\rb \lsqb 1 +  \gamma_1\frac{P_J}{P_1 - \gamma_1 P_2} 
  \lb\frac{d_2}{d_3}\rb^\alpha\rsqb^{-1}\rcb. \label{eq:fdasnoise6}
\end{align}
From (\ref{eq:fdasnoise6}), we can see that the event $\mathcal{D}_{1/1,2}^s$ can occur with a non-zero probability if
\begin{align}
 \frac{P_1}{P_2} > \gamma_1. \label{eq:fdasnoise6a}
\end{align}
Receiver~$2$ can decode the packet sent by the second queue when the following event is true.
\begin{align}
\mathcal{D}_{2/1,2} = \lcb \frac{P_2 \htwosq d_2^{-\alpha}}{1 + P_1 \htwosq d_2^{-\alpha} + 
P_J \honetwosq d_3^{-\alpha}} \geq \gamma_2\rcb.  \label{eq:fdasnoise7}
\end{align}
The success probability in this case is
\begin{align}
& \prob \lb \mathcal{D}_{2/1,2}\rb \nonumber \\
&  = \exp\lb - \frac{\gamma_2 d_2^\alpha}{P_2 - \gamma_2 P_1}\rb \lsqb 1 + 
\gamma_2 \frac{P_J}{P_2 - \gamma_2 P_1} \lb \frac{d_2}{d_3}\rb^{\alpha}\rsqb^{-1}.  \label{eq:fdasnoise8}
\end{align}
The event $\mathcal{D}_{2/1,2}$ can occur with nonzero probability if the following condition is satisfied
\begin{align}
 \frac{P_2}{P_1} > \gamma_2. \label{eq:fdasnoise8a}
\end{align}
Hence, both the events $\mathcal{D}_{1/1,2}^s$ and $\mathcal{D}_{2/1,2}$ can occur, provided the conditions in (\ref{eq:fdasnoise6a}) and (\ref{eq:fdasnoise8a}) are satisfied. 

Using (\ref{eq:fdasnoise5}), (\ref{eq:fdasnoise6}) and (\ref{eq:fdasnoise8}), the stability region in (\ref{eq:stability-region1}) becomes (\ref{eq:fdasnoise9}). Using (\ref{eq:fdasnoise2}), (\ref{eq:fdasnoise6}) and (\ref{eq:fdasnoise8}), the stability region in (\ref{eq:stability-region2}) becomes (\ref{eq:fdasnoise10}). Finally, the stability region of the system with the secrecy constraint is obtained by taking union of $\mathcal{R}_1$ in (\ref{eq:fdasnoise9}) and $\mathcal{R}_2$ in (\ref{eq:fdasnoise10}).

\begin{figure*}
\begin{align}
\mathcal{R}_1  = &\lcb (\lambda_1, \lambda_2): \frac{1 - \Ronetermone \frac{\Ronetermtwo}{\Ronetermfour}}{\Ronetermsix} \lambda_2\nonumber \right. \\
& +  \left. \frac{\lambda_1}{\exp(-\frac{\gamma_1 d_1^\alpha}{P_1}(1 + g^2 P_J)) \lcb \Ronetermtwo \rcb} < 1, 
\lambda_2 < \frac{\exp\lb - \frac{\gamma_2 d_2^\alpha}{P_2 - \gamma_2 P_1}\rb}{\Ronetermseven} \rcb. 
 \label{eq:fdasnoise9} \\ 
\mathcal{R}_2   =& \lcb (\lambda_1, \lambda_2):  \frac{1 - \exp\lb - \lb \frac{1}{P_2 - \gamma_2 P_1} - \frac{1}{P_2}\rb\gamma_2 d_2^\alpha\rb \lb \frac{\Rtwotermone}{\Rtwotermtwo}\rb^{-1}}{\exp \lb - \frac{\gamma_1 d_1^\alpha}{P_1 - \gamma_1 P_2} (1 + g^2 P_J)\rb\lcb \Rtwotermfour\rcb} \lambda_1  \right. \nonumber \\
& + \left. \frac{\lambda_2}{\exp \lb  - \frac{\gamma_2 d_2^\alpha}{P_2}\rb \lsqb \Rtwotermtwo 
\rsqb^{-1}} < 1, \right. \nonumber \\
& \left. \lambda_1 \leq \exp \lb - \frac{\gamma_1 d_1^\alpha}{P_1 - \gamma_1 P_2} (1 + g^2 P_J)\rb\lcb \Rtwotermfour\rcb 
\rcb.  \label{eq:fdasnoise10} \\ \nonumber \\ \hline \nonumber
\end{align}
\end{figure*}

\subsection{Receiver~$1$ Performs Successive Decoding and Receiver~$2$ Treats Interference as Noise}\label{sec:rx1-SD-rx2-TIN}
In this subsection, we consider the case, where receiver~$1$ performs successive decoding \cite{gamal-book-2011}, i.e., it first decodes the packet intended to receiver~$2$, and cancels its effect from the output. Then, receiver~$1$ tries to decode its intended packet. The second receiver cannot perform successive decoding, and hence treats other user's packet as noise while decoding its intended packet. When $(Q_1 = 0, Q_2 \neq 0)$ and $(Q_1 \neq 0, Q_2 = 0)$, the success probability for receiver~$2$~and~$1$ are given by (\ref{eq:fdasnoise2}) and (\ref{eq:fdasnoise5}), respectively. The expression for success probability changes only for receiver~$1$ in the case of $(Q_1 \neq 0, Q_2 \neq 0)$, which is obtained as follows.

Receiver~$1$ can perform successive decoding under the secrecy constraint when the following event is true.
\begin{align}
\mathcal{D}_{1/2,1}^{s} & \!=\! \lcb \frac{P_2 \honesq d_1^{-\alpha}}{1 + P_1 \honesq d_1^{-\alpha} + g^2 P_J} \!\geq\! \gamma_2,
 \frac{P_1 \honesq d_1^{-\alpha}}{1 + g^2 P_J} \geq \gamma_1, \right. \nonumber \\
 & \qquad \left. \frac{P_1 \htwosq d_2^{-\alpha}}{1 + P_2 \htwosq d_2^{-\alpha} + \honetwosq d_3^{-\alpha} P_J} < \gamma_1\rcb. \label{eq:succdecode1}  
\end{align} 
The success probability of receiver~$1$ is then given by

{%\scriptsize
\begin{align}
& \prob(\mathcal{D}_{1/2,1}^{s}) \nonumber \\
& = \prob\lb \honesq \geq \max\lcb \frac{\gamma_2 d_1^{\alpha}}{P_2 - \gamma_2 P_1}, \frac{\gamma_1 d_1^{\alpha}}{P_1} \rcb\rb \times \nonumber \\
& \qquad \prob\lb \htwosq \leq \frac{\gamma_1 d_2^{\alpha}}{P_1 - \gamma_1 P_2}\rb
 \nonumber \\
 %New eqn
 & = \exp\lb - \!\!\max\lcb \frac{\gamma_2 \lb 1 + g^2 P_J\rb d_1^{\alpha}}{P_2 - \gamma_2 P_1}, \frac{\gamma_1 \lb 1 + g^2 P_J\rb d_1^{\alpha}}{P_1} \rcb\rb \times
  \nonumber \\
 &\lsqb\! 1 \!-\! \exp\lb -\frac{\gamma_1 d_2^{\alpha}}{P_1 - \gamma_1 P_2}\rb  \lcb 1 \!+\! \gamma_1 \frac{P_J}{P_1 - \gamma_1 P_2} \lb \frac{d_2}{d_3}\rb^\alpha\rcb^{-1}\rsqb. \label{eq:succdecode2}  
\end{align}}
The success probability of receiver~$2$ is given by (\ref{eq:fdasnoise8}).

Replacing the success probabilities obtained for different cases in (\ref{eq:stability-region1}) 
and (\ref{eq:stability-region2}), the stability region for the successive decoding scheme can be determined.

\subsection{Receiver~$1$ Treats Interference as Noise and Receiver~$2$ Performs Successive Decoding}\label{sec:stability-region-rx2}
In this section, we consider the case where receiver~$1$ cannot perform successive decoding. Hence, it decodes its own packet by treating interference as noise. It is assumed that receiver~$2$ can perform successive decoding. In this case, receiver~$2$ decodes the packet of the first queue and then, tries to decode its intended packet, i.e., the packet sent from the second queue. If receiver~$2$ fails in successive decoding, then it attempts to decode its intended packet by treating interference as noise. The jamming signal sent by receiver~$1$ creates additional noise at receiver~$2$ and hence, it can assist receiver~$1$ to achieve nonzero secrecy throughput. However, the jamming signal transmitted by receiver~$1$ creates self-interference at receiver~$1$ and the strength of the interference depends on the degree of self-interference cancelation. Receiver~$1$, while decoding its own packet, treats the packet intended for receiver~$2$ as well as the residual self-interference cancelation from the jamming signal as noise. In the following, the success probability for both the receivers are obtained for different status of the queues.

When $Q_1 = 0$ and $Q_2 \neq 0$, and $Q_1 \neq 0$ and $Q_2 = 0$, the success probabilities $\prob(\mathcal{D}_{2/2})$ and $\prob(\mathcal{D}_{1/1}^s)$ are same as in (\ref{eq:fdasnoise2}) and (\ref{eq:fdasnoise5}), respectively. Hence, it is required to consider the following case only. 

\subsubsection{When $Q_1 \neq 0$ and $Q_2 \neq 0$} 
In this case we assume that receiver $2$ can perform successive decoding. For the event $\mathcal{D}_{1/1,2}^s$, receiver~$1$ should be able to decode its own packet as well as receiver~$2$ should not be able to decode the packet intended to receiver~$1$ either by treating interference as noise or successive decoding. In the case of successive decoding, receiver~$2$ tries to decode its intended packet first and then tries to cancel its effect to decode the packet sent from queue~$1$. However, in both cases, the jamming signal sent from receiver~$1$ remains as noise at receiver~$2$. 

The probability that receiver~$1$ decodes its intended packet with the secrecy constraint is determined as
\begin{align}
 \prob \lb \mathcal{D}_{1/1,2}^s \rb & =  \prob \lb \mathcal{D}_{1/1,2}^{s, \text{2SD}} \rb + \prob \lb \mathcal{D}_{1/1,2}^{s, \text{2IAN}} \rb - \nonumber \\
& \qquad \prob \lb \mathcal{D}_{1/1,2}^{s, \text{2SD}} \cap \mathcal{D}_{1/1,2}^{s, \text{2IAN}} \rb, \label{eq:fdsuccrxtwo6a} 
\end{align}
where $\mathcal{D}_{1/1,2}^{s, \text{2SD}}$ denotes the event that receiver~$1$ receives secretly its information if receiver~$2$ can perform successive decoding (SD), i.e., receiver~$2$ decodes its intended packet first, then tries to decode the packet intended to receiver~$1$, and $\mathcal{D}_{1/1,2}^{s, \text{2IAN}}$ denotes the event that receiver~$1$ can decode its information secretly if receiver~$2$ performs treating interference as noise (IAN).

The probability of the event $\mathcal{D}_{1/1,2}^{s, \text{2SD}}$ is calculated as follows
\begin{align}
\prob\lb  \mathcal{D}_{1/1,2}^{s, \text{2SD}} \rb & = \prob\lb A_1 \cap A_2 \cap A_3\rb \nonumber \\
%New eqnarray
& = \prob \lb A_1|A_2 \cap A_3\rb \prob \lb A_2 \cap A_3\rb \nonumber \\
%New eqnarray
& = \prob \lb A_1 \rb \prob \lb A_2 \cap A_3\rb, \label{eq:fdsuccrxtwo6} 
\end{align}
where the events $A_1$, $A_2$ and $A_3$ are defined as
\begin{align}
& A_1 = \lcb \frac{P_1 \honesq d_1^{-\alpha}}{1 + P_2 \honesq d_1^{-\alpha} + P_J \hself} \geq \gamma_1\rcb,  \nonumber \\
& A_2 = \lcb \frac{P_2 \htwosq d_2^{-\alpha}}{1 + P_1 \htwosq d_2^{-\alpha} + P_J \honetwosq d_3^{-\alpha}} \geq \gamma_2 \rcb, \nonumber \\
%New line
& A_3 = \lcb \frac{P_1 \htwosq d_2^{-\alpha}}{1 + P_J \honetwosq d_3^{-\alpha}} < \gamma_1 \rcb.
\end{align}
Note that in the above, $A_1$ is independent of $A_2$ and $A_3$. Now, we consider the evaluation of the following term in (\ref{eq:fdsuccrxtwo6})
\begin{align}
\prob\lb A_1\rb & = \prob \lcb \frac{P_1 \honesq d_1^{-\alpha}}{1 + P_2 \honesq d_1^{-\alpha} + P_J \hself} \geq \gamma_1\rcb \nonumber \\
%Ne eqnarray
& = \exp\lb - \frac{\gamma_1 d_1^\alpha}{P_1 - \gamma_1 P_2} \lb 1 + P_J \hself\rb\rb. \label{eq:fdsuccrxtwo7} 
\end{align}
Consider the evaluation of the second term in (\ref{eq:fdsuccrxtwo6})

%\scriptsize
\begin{align}
& \prob\lb A_2 \cap A_3\rb \nonumber \\
%New eqnarray
& = \prob \lcb \frac{P_2 \htwosq d_2^{-\alpha}}{1 + P_1 \htwosq d_2^{-\alpha} + P_J \honetwosq d_3^{-\alpha}} \geq \gamma_2, \right. \nonumber \\
& \qquad \qquad \left. \frac{P_1 \htwosq d_2^{-\alpha}}{1 + P_J \honetwosq d_3^{-\alpha}} < \gamma_1 \rcb. \label{eq:fdsuccrxtwo8} 
\end{align}
For the event $\{A_2 \cap A_3\}$ to be feasible, the following condition is required to be satisfied
\begin{align}
\frac{P_1}{P_2} < \frac{\gamma_1}{(1 + \gamma_1)\gamma_2}. \label{eq:fdsuccrxtwo9} 
\end{align}
On further simplification, $\prob\lb A_2 \cap A_3\rb$ becomes
\begin{align}
 & \prob\lb A_2 \cap A_3\rb \nonumber \\
 %New line
 & \!= \prob \lb \frac{\gamma_2 (1 + P_J \honetwosq d_3^{-\alpha})}{\lb P_2 - \gamma_2 P_1\rb d_2^{-\alpha}} \leq \htwosq \!\leq\!  \right. \nonumber \\
 & \qquad \left. \frac{\gamma_1 (1 + P_J \honetwosq d_3^{-\alpha})}{ P_1 d_2^{-\alpha}}\rb \nonumber \\
%New line
& = \int_0^{\infty} \!\prob \lb \frac{\gamma_2 (1+P_J \honetwosq d_3^{-\alpha} )}{(P_2 - \gamma_2 P_1)d_2^{-\alpha}} \leq \htwosq  \right. \nonumber \\
& \qquad \quad \left. < \frac{\gamma_1 (1+ P_J \honetwosq d_3^{-\alpha})}{P_1 d_2^{-\alpha}} \bigg | \honetwosq=x \rb f_{\honetwosq}(x)  \,  \mathrm{d}x \nonumber \\
%New eqnarray
& = \frac{\exp\lb -\frac{\gamma_2 d_2^\alpha}{P_2 - \gamma_2 P_1}\rb}{1 + \frac{\gamma_2 P_J}{P_2 - \gamma_2 P_1}\lb \frac{d_2}{d_3}\rb^{\alpha}} - \frac{\exp \lb -\frac{\gamma_1 d_2^\alpha}{P_1} \rb}{1 + \gamma_1 \frac{P_J}{P_1} \lb \frac{d_2}{d_3}\rb^\alpha}, \label{eq:fdsuccrxtwo10}
\end{align}
where for the above integral to exist the following condition must be satisfied 
\begin{align}
& \lb \frac{d_2}{d_3}\rb^{\alpha} > \frac{\gamma_2 P_1 - P_2}{\gamma_2 P_J}. \label{eq:fdsuccrxtwo11}
\end{align}

Using (\ref{eq:fdsuccrxtwo7}) and (\ref{eq:fdsuccrxtwo10}), (\ref{eq:fdsuccrxtwo6}) becomes 
\begin{align}
& \prob\lb  \mathcal{D}_{1/1,2}^{s, \text{2SD}} \rb = \exp\lb - \frac{\gamma_1 d_1^\alpha}{P_1 - \gamma_1 P_2} \lb 1 + P_J \hself\rb\rb \times \nonumber \\
& \qquad \lsqb \frac{\exp\lb \frac{-\gamma_2 d_2^\alpha}{P_2 - \gamma_2 P_1}\rb}{1 + \frac{\gamma_2 P_J}{P_2 - \gamma_2 P_1}\lb \frac{d_2}{d_3}\rb^{\alpha}} - \frac{\exp \lb \frac{-\gamma_1 d_2^\alpha}{P_1} \rb}{1 + \gamma_1 \frac{P_J}{P_1} \lb \frac{d_2}{d_3}\rb^\alpha}\rsqb. \label{eq:fdsuccrxtwo12}
\end{align}

Now, consider the evaluation of $\prob\lb \mathcal{D}_{1/1,2}^{s, \text{2IAN}}\rb$ in (\ref{eq:fdsuccrxtwo6a})
\begin{align}
 \prob\lb \mathcal{D}_{1/1,2}^{s, \text{2IAN}}\rb & = \prob \lb A_1\rb \prob \lb A_4 \cap A_5\rb \nonumber \\
 %New line
= &  \exp\lb - \frac{\gamma_1 d_1^\alpha}{P_1 - \gamma_1 P_2} \lb 1 + P_J \hself\rb\rb \prob \lb A_4 \cap A_5\rb, \label{eq:fdsuccrxtwo13}
\end{align}
where the events $A_4$ and $A_5$ are defined as
\begin{align}
 & A_4 = \lcb \frac{P_2 \htwosq d_2^{-\alpha}}{1 + P_1 \htwosq d_2^{-\alpha} + P_J \honetwosq d_3^{-\alpha}} < \gamma_2\rcb, \text{ and } \nonumber \\
 & A_5 = \lcb \frac{P_1 \htwosq d_2^{-\alpha}}{1 + P_2 \htwosq d_2^{-\alpha} + P_J \honetwosq d_3^{-\alpha}} < \gamma_1\rcb. \label{eq:fdsuccrxtwo14}
\end{align}
Consider the following
\begin{align}
& \prob\lb A_4 \cap A_5 \rb \nonumber \\
%New line
& = \prob\lcb \frac{P_2 \htwosq d_2^{-\alpha}}{1 + P_1 \htwosq d_2^{-\alpha} + P_J \honetwosq d_3^{-\alpha}} < \gamma_2, \right. \nonumber \\
& \qquad \qquad \left. \frac{P_1 \htwosq d_2^{-\alpha}}{1 + P_2 \htwosq d_2^{-\alpha} + P_J \honetwosq d_3^{-\alpha}} < \gamma_1 \rcb \nonumber \\
%New line
& =  \prob \lcb \frac{\htwosq}{1 + P_J \honetwosq d_3^{-\alpha}} < \frac{d_2^\alpha}{\max\lcb \frac{P_2}{\gamma_2} - P_1, \frac{P_1}{\gamma_1} - P_2\rcb} \rcb \nonumber \\
%New line
& = 1 - \exp\lb - \frac{d_2^\alpha}{\max\lcb \frac{P_2}{\gamma_2} - P_1, \frac{P_1}{\gamma_1} - P_2\rcb} \rb \times \nonumber \\
& \qquad \qquad  \lcb 1 + \frac{d_2^\alpha}{\max\lcb \frac{P_2}{\gamma_2} - P_1, \frac{P_1}{\gamma_1} - P_2\rcb} P_J d_3^{-\alpha}\rcb^{-1}.
\label{eq:fdsuccrxtwo15}
\end{align}

Using (\ref{eq:fdsuccrxtwo15}), (\ref{eq:fdsuccrxtwo13}) becomes
\begin{align}
& \prob\lb \mathcal{D}_{1/1,2}^{s, \text{2IAN}}\rb \nonumber \\
& = \exp\lb - \frac{\gamma_1 d_1^\alpha}{P_1 - \gamma_1 P_2} \lb 1 + P_J \hself\rb\rb \times \nonumber \\
& \lb 1 - \exp\lb - \frac{d_2^\alpha}{\max\lcb \frac{P_2}{\gamma_2} - P_1, \frac{P_1}{\gamma_1} - P_2\rcb} \rb \times \right. \nonumber \\
& \left.  \lcb 1 + \frac{d_2^\alpha}{\max\lcb \frac{P_2}{\gamma_2} - P_1, \frac{P_1}{\gamma_1} - P_2\rcb} P_J d_3^{-\alpha}\rcb^{-1} \rb. \label{eq:fdsuccrxtwo16}
\end{align}

It can be noticed that as the events $A_2$ and $A_4$ are mutually exclusive of each other, we have 
\begin{align}
\prob \lb \mathcal{D}_{1/1,2}^{s, \text{2SD}} \cap \mathcal{D}_{1/1,2}^{s, \text{2IAN}} \rb = 0. \label{eq:fdsuccrxtwo16b}
\end{align}

Using (\ref{eq:fdsuccrxtwo12}), (\ref{eq:fdsuccrxtwo16}), and (\ref{eq:fdsuccrxtwo16b}), (\ref{eq:fdsuccrxtwo6a}) becomes
\begin{align}
 & \prob \lb \mathcal{D}_{1/1,2}^s \rb \nonumber \\
 & = \exp\lb - \frac{\gamma_1 d_1^\alpha}{P_1 - \gamma_1 P_2} \lb 1 + P_J \hself\rb\rb \lsqb \frac{\exp\lb \frac{-\gamma_2 d_2^\alpha}{P_2 - \gamma_2 P_1}\rb}{1 + \frac{\gamma_2 P_J}{P_2 - \gamma_2 P_1}\lb \frac{d_2}{d_3}\rb^{\alpha}} - \right. \nonumber \\
 &\left. \frac{\exp \lb \frac{-\gamma_1 d_2^\alpha}{P_1} \rb}{1 + \gamma_1 \frac{P_J}{P_1} \lb \frac{d_2}{d_3}\rb^\alpha}\rsqb  + \exp\lb - \frac{\gamma_1 d_1^\alpha}{P_1 - \gamma_1 P_2} \lb 1 + P_J \hself\rb\rb \times \nonumber \\
& \lb 1 - \exp\lb - \frac{d_2^\alpha}{\max\lcb \frac{P_2}{\gamma_2} - P_1, \frac{P_1}{\gamma_1} - P_2\rcb} \rb \times \right. \nonumber \\
& \left. \lcb 1 + \frac{d_2^\alpha}{\max\lcb \frac{P_2}{\gamma_2} - P_1, \frac{P_1}{\gamma_1} - P_2\rcb} P_J d_3^{-\alpha}\rcb^{-1} \rb. \label{eq:fdsuccrxtwo16a}
\end{align}

In the following, the success probability for receiver~$2$ is obtained. In this case, receiver~$2$ first attempts to decode its intended packet by successive decoding. If it fails in successive decoding, it decodes by treating interference as noise.
\begin{align}
\prob\lb \mathcal{D}_{2|1,2} \rb & = \prob \lb \mathcal{D}_{2|1,2}^{2\text{IAN}} \cup \mathcal{D}_{2|1,2}^{2\text{SD}}\rb \nonumber \\
%New line
& =\! \prob \lb \mathcal{D}_{2|1,2}^{2\text{IAN}}\rb \!+\! \prob \lb \mathcal{D}_{2|1,2}^{2\text{SD}}\rb \!-\! \prob\lb \mathcal{D}_{2|1,2}^{2\text{IAN}} \cap  \mathcal{D}_{2|1,2}^{2\text{SD}} \rb, 
\label{eq:fdsuccrxtwo17}
\end{align}
where $\mathcal{D}_{2|1,2}^{2\text{IAN}}$ denotes the event that receiver~$2$ is able to decode its intended packet with treating interference as noise and $\mathcal{D}_{2|1,2}^{2\text{SD}}$ denotes the event that receiver~$2$ is able to decode its intended packet with successive decoding, i.e., receiver~$2$ decodes the packet intended to receiver~$1$, then attempts to decode its own packet. To calculate $\prob\lb \mathcal{D}_{2|1,2} \rb$, consider the following event
\begin{align}
& B_1 = \lcb \frac{P_2 \htwosq d_2^{-\alpha}}{1 + P_1 \htwosq d_2^{-\alpha} + P_J \honetwosq d_3^{-\alpha}} \geq \gamma_2\rcb, \nonumber \\
&B_2 = \lcb \frac{P_1 \htwosq d_2^{-\alpha}}{1 + P_2 \htwosq d_2^{-\alpha} + P_J \honetwosq d_3^{-\alpha}} \geq \gamma_1\rcb, \nonumber \\
%New line
& B_3 = \lcb \frac{P_2 \htwosq d_2^{-\alpha}}{1 + P_J \honetwosq d_3^{-\alpha}} \geq \gamma_2 \rcb \label{eq:fdsuccrxtwo17a}
\end{align}

Now, consider the following
\begin{align}
& \prob\lb \mathcal{D}_{2|1,2}^{2\text{IAN}} \rb = \prob \lb B_1\rb \nonumber \\
%New line
& = \exp\lb - \frac{\gamma_2 d_2^\alpha}{P_2 - \gamma_2 P_1}\rb \lsqb 1 + \gamma_2 \frac{P_J}{P_2 - \gamma_2 P_1} \lb\frac{d_2}{d_3}\rb^{\alpha}\rsqb^{-1}. \label{eq:fdsuccrxtwo18}
\end{align}
The equation above is obtained using the results derived in Section~\ref{sec:decoding-cons-rx2} (See (\ref{eq:fdasnoise7}) and (\ref{eq:fdasnoise8})).

Consider the calculation of probability of the event $\mathcal{D}_{2|1,2}^{2\text{SD}} = B_2 \cap B_3 \cap B_1^c$. Since the events $\mathcal{D}_{2|1,2}^{2\text{IAN}}$ and $\mathcal{D}_{2|1,2}^{2\text{SD}}$ are mutually exclusive, we have
\begin{align}
& \prob \lb \mathcal{D}_{2|1,2}^{2\text{SD}} \rb  = \prob \lb B_2 \cap B_3 \cap B_1^c\rb \nonumber \\
%New line
& = \prob \lb \frac{P_1 \htwosq d_2^{-\alpha}}{1 + P_2 \htwosq d_2^{-\alpha} + P_J \honetwosq d_3^{-\alpha}} \geq \gamma_1, \right. \nonumber \\
& \qquad \left. \frac{P_2 \htwosq d_2^{-\alpha}}{1 + P_J \honetwosq d_3^{-\alpha}} \geq \gamma_2, \right. \nonumber \\
& \qquad \left. \frac{P_2 \htwosq d_2^{-\alpha}}{1 + P_1 \htwosq d_2^{-\alpha} + P_J \honetwosq d_3^{-\alpha}} < \gamma_2\rb \nonumber \\
%New line
& = \prob\lb \gamma_2' \leq \frac{\htwosq}{1 + P_J \honetwosq d_3^{-\alpha}} \leq \gamma_2'' \rb,  \label{eq:fdsuccrxtwo19}
%New line
\end{align}
where $\gamma_2' \triangleq \min\lcb \frac{\gamma_1 d_2^{\alpha}}{P_1 - \gamma_1 P_2}, \frac{\gamma_2 d_2^\alpha}{P_2}\rcb$ and $\gamma_2'' = \frac{\gamma_2 d_2^\alpha}{P_2 - \gamma_2 P_1}$. For this probability to exist, following condition is required to be satisfied
\begin{align}
%& \gamma_2' < \gamma_2'', \nonumber \\
& \frac{\gamma_1}{P_1 - \gamma_1 P_2} < \frac{\gamma_2}{P_2 - \gamma_2 P_1}\label{eq:fdsuccrxtwo20}
\end{align}

When (\ref{eq:fdsuccrxtwo20}) is satisfied, (\ref{eq:fdsuccrxtwo19}) becomes
\begin{align}
& \prob \lb \mathcal{D}_{2|1,2}^{2\text{SD}} \rb  \nonumber \\
%New line
& = \prob \lb \gamma_2' \lb 1 + P_J \honetwosq d_3^{-\alpha}\rb \leq \htwosq \leq \right. \nonumber \\
& \left. \gamma_2'' \lb 1 + P_J \honetwosq d_3^{-\alpha}\rb\rb \nonumber \\
%New line
& = \int_0^{\infty} \!\prob \lb \gamma_2'(1+P_J \honetwosq d_3^{-\alpha}) \leq \htwosq  \right. \nonumber \\
& \qquad \qquad \left. < \gamma_2''(1+ P_J \honetwosq d_3^{-\alpha}) \bigg | \honetwosq=x \rb f_{\honetwosq}(x)  \,  \mathrm{d}x \nonumber \\
%New eqnarray
& = \frac{\exp\lb- \gamma_2'\rb}{1 + \gamma_2' P_J d_3^{-\alpha}}- \frac{\exp\lb- \gamma_2''\rb}{1 + \gamma_2'' P_J d_3^{-\alpha}}. \label{eq:fdsuccrxtwo21}
\end{align}
Using (\ref{eq:fdsuccrxtwo18}) and (\ref{eq:fdsuccrxtwo21}) and noticing that $\prob\lb \mathcal{D}_{2|1,2}^{2\text{IAN}} \cap  \mathcal{D}_{2|1,2}^{2\text{SD}} \rb = 0$, (\ref{eq:fdsuccrxtwo17}) becomes 
\begin{align}
& \prob\lb \mathcal{D}_{2|1,2} \rb \nonumber \\
& = \exp\lb - \frac{\gamma_2 d_2^\alpha}{P_2 - \gamma_2 P_1}\rb \lsqb 1 + \gamma_2 \frac{P_J}{P_2 - \gamma_2 P_1} \lb\frac{d_2}{d_3}\rb^{\alpha}\rsqb^{-1} \nonumber \\
& \qquad \qquad \qquad  + \frac{\exp\lb- \gamma_2'\rb}{1 + \gamma_2' P_J d_3^{-\alpha}}- \frac{\exp\lb -\gamma_2''\rb}{1 + \gamma_2'' P_J d_3^{-\alpha}}. \label{eq:fdsuccrxtwo22}
\end{align}

Replacing the success probabilities obtained for different cases in (\ref{eq:stability-region1}) 
and (\ref{eq:stability-region2}), the stability region can be determined.

\subsection{Remarks}
\begin{enumerate}
\item The stability region for the BC with the secrecy constraint when receiver~$1$ does not have full-duplex capability can be obtained as a special case by setting $P_J=0$ for the different decoding schemes.
\item In the third case (Section~\ref{sec:stability-region-rx2}), if receiver~$2$ can only perform successive decoding (i.e., it cannot treat interference as noise if it fails in successive decoding), then the success probabilities $\prob \lb \mathcal{D}_{1/1,2}^s \rb$ and $\prob\lb \mathcal{D}_{2|1,2} \rb$ are obtained by setting $ \prob \lb \mathcal{D}_{1/1,2}^{s, \text{2IAN}} \rb = 0$ and $\prob \lb \mathcal{D}_{2|1,2}^{2\text{IAN}} \rb=0 $ in (\ref{eq:fdsuccrxtwo6a}) and (\ref{eq:fdsuccrxtwo17}), respectively.
\item The results derived in this section are also useful for other communication scenarios such as the cases with different kind of data arrivals and congestion levels at the queues. We can also apply the results for the optimal design of the system. For example, the jamming power of receiver 1 can be optimized to enhance the system throughput as we present in Section~\ref{sec:optimization}.

\item Using the result in \cite{gamal-book-2011}, it is not difficult to show that both the receivers cannot perform successive decoding simultaneously since it results in infeasible power allocation even for the BC without the secrecy constraint as given by
\begin{align}
P_1 \leq \frac{|h_1|^2 - |h_2|^2}{|h_1|^2 |h_2|^2} \quad \text{and} \quad P_2 \leq \frac{|h_2|^2 - |h_1|^2}{|h_1|^2 |h_2|^2}. \label{eq:power} 
\end{align}
From (\ref{eq:power}), we can see that $P_1$ and $P_2$ cannot be positive simultaneously. Note that the stability region of the BC without secrecy constraint $(\mathcal{R}^{\text{ws}})$ will include the stability region with the secrecy constraint $(\mathcal{R})$, i.e., $\mathcal{R} \subseteq \mathcal{R}^{\text{ws}}$ and hence, we have not considered the case of both the receivers performing successive decoding simultaneously.
\end{enumerate}

\section{Numerical Results}\label{sec:numerical-result}
In this section, numerical results are presented first for the stability region for different decoding schemes at the receivers. Then, the closures of the stability regions are obtained for the different cases. Finally, numerical results are presented for the optimization problem of jamming power considered in Section~\ref{sec:optimization}.
\subsection{Stability Region Analysis with Different Decoding Schemes}
In this subsection, we present the stability region for difference decoding capabilities at the receivers. In all figures, $(\text{DC}_1, \text{DC}_2)$ means the decoding scheme at receiver $i$ is $\text{DC}_i$, which is either the decoding by treating interference as noise (TIN) or the successive decoding (SD).

\subsubsection{No full-duplex capability at receiver 1}
\begin{figure}[t] 
	\centering
	\includegraphics[width=3.4in]{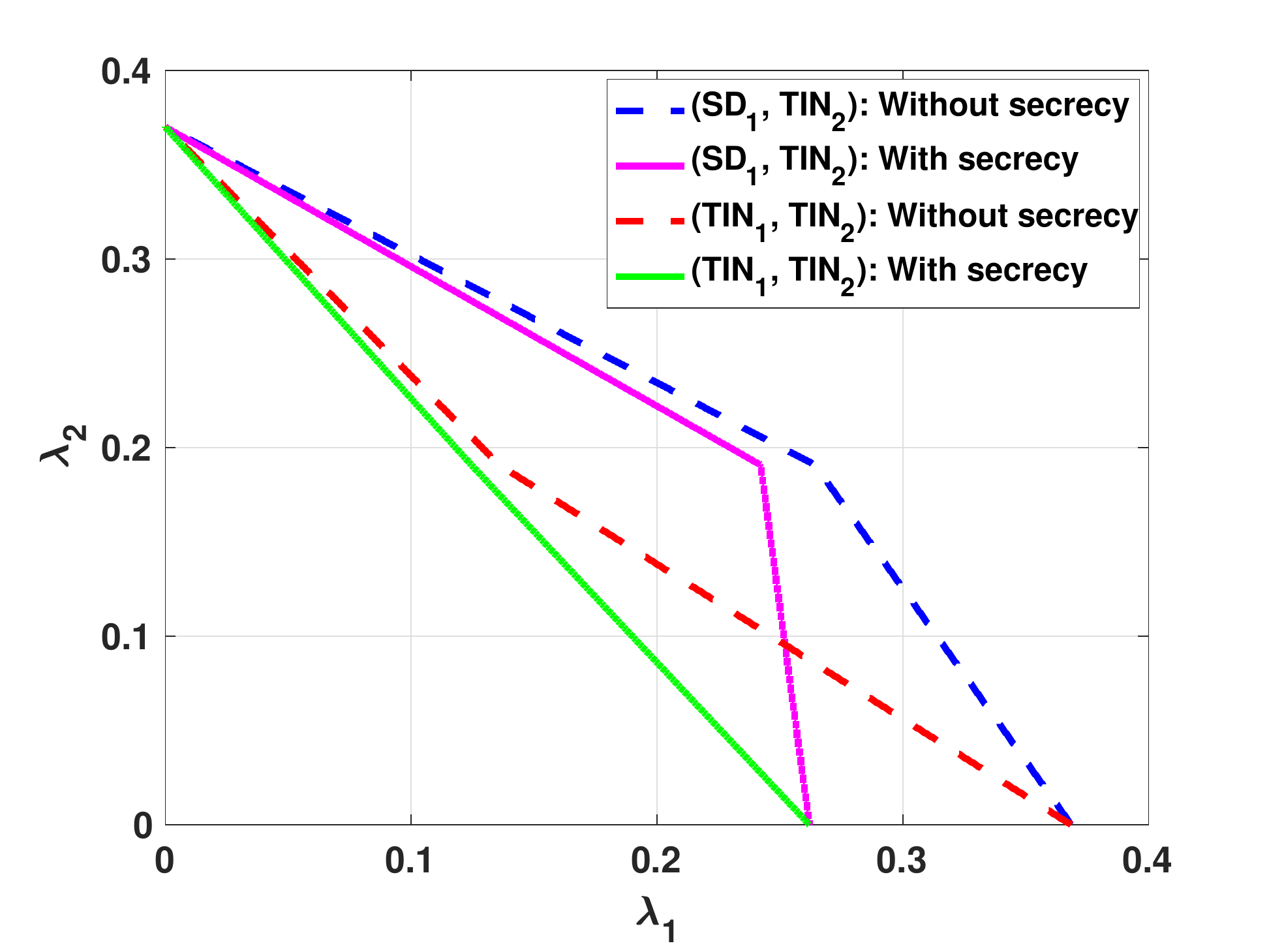}
	\caption{The stability region where $P_1= P_2 = 100$, $P_J= 0$, $\gamma_1=0.5$, $\gamma_2 = 0.4$, $d=10$, $d_1 = d$, $d_2 = 1.1d$, and $\alpha= 2.3$.}\label{fig:result1}
	\vspace{-0.3cm}
\end{figure}
The stability regions are plotted in Fig.~\ref{fig:result1} when receiver~$1$ does not send any jamming signal to receiver~$2$, i.e., $P_j = 0$, and receiver 2 decodes by treating interference as noise (TIN) due to a decoding constraint. In this figure, different decoding capabilities for receiver 1 (i.e., SD and TIN) are considered for both cases with/without the secrecy constraint at receiver 1. Note that the stability regions without the secrecy constraint are plotted based on \cite{pappas-iccBC-2016} for both the decoding schemes of receiver 1. The stability regions with the secrecy constraints are plotted using the results derived in Sections~\ref{sec:decoding-cons-rx2} and \ref{sec:rx1-SD-rx2-TIN}.

In Fig.~\ref{fig:result1}, the same transmit power for packet in the queues are assumed as $P_1 = P_2 = 100$. From Fig.~\ref{fig:result1}, it can be noticed that due to the secrecy constraint for packets intended to receiver~$1$, the stability region has reduced compared to the case when there is no secrecy constraint at receiver~$2$. One can also notice that as the value of $\lambda_1$ decreases, the penalty on the stability region due to the secrecy constraint also decreases. Expectedly, when there is no packet for receiver~$1$, there is no penalty on the throughput for receiver~$2$ as there is no secrecy constraint for the packet intended to receiver~$2$. It is interesting to note that the stability region with the secrecy constraint for the successive decoding can be larger than the stability region without the secrecy constraint for the TIN case, for some values of $\lambda_1$ and $\lambda_2$. 

\subsubsection{With full-duplex capability at receiver 1}
\begin{figure}[t]
	\centering
	\includegraphics[width=3.4in]{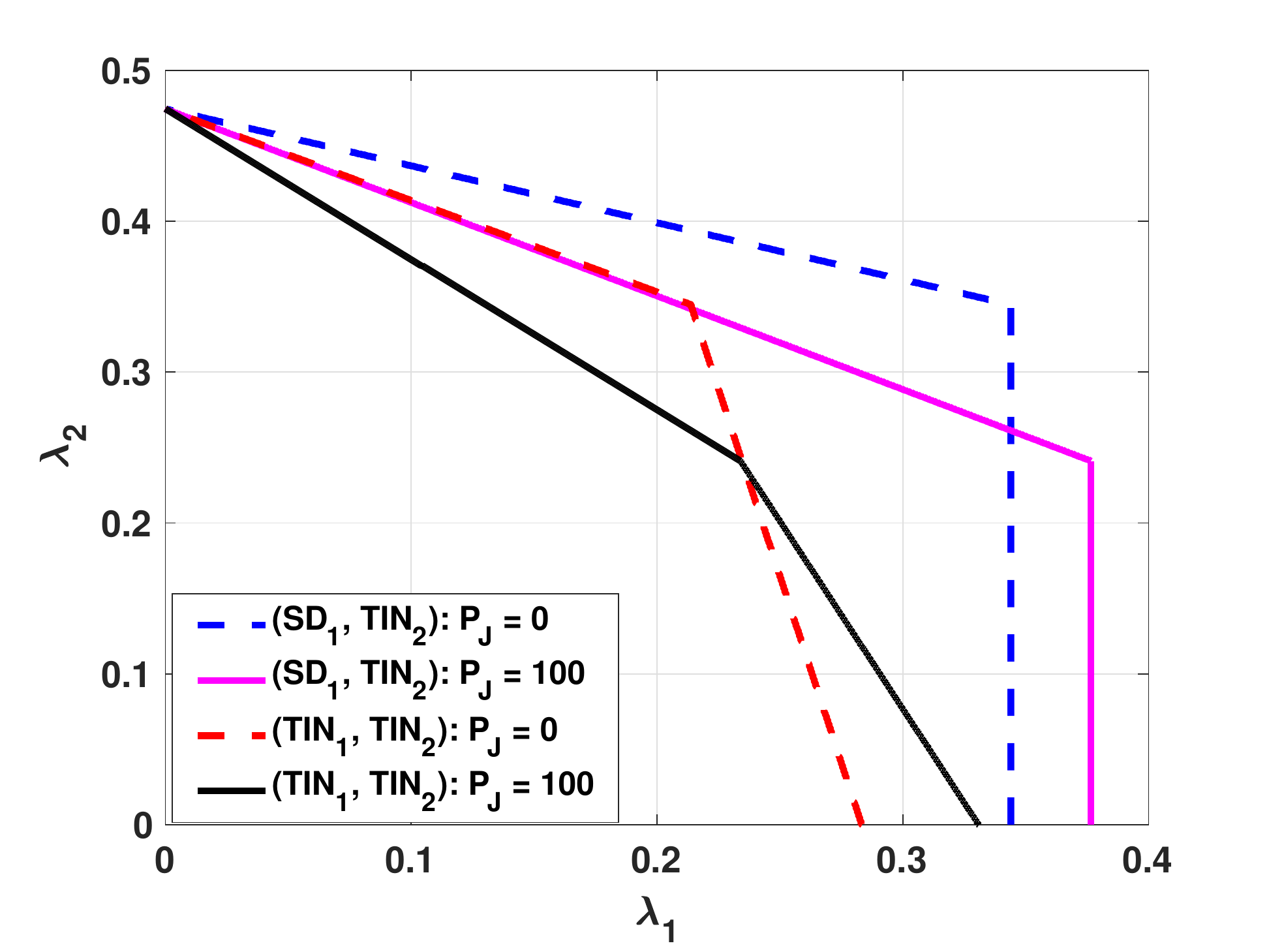}
	\caption{The stability region with and without full-duplex capability at receiver~$1$: $P_1= 100, P_2 = 100$, $\gamma_1=0.4$, $\gamma_2 = 0.3$, $d=10$, $d_1 = d$, $d_2 = d_3 = 1.1d$, $g = 10^{-3}$, and $\alpha= 2.3$.}\label{fig:result3}
	\vspace{-0.4cm}
\end{figure}

In this case, the stability region is depicted when receiver~$1$ sends a jamming signal to receiver~$2$ and it is assumed that receiver~$2$ cannot perform successive decoding. In Fig.~\ref{fig:result3}, the stability regions are plotted using the results derived in Sections~\ref{sec:decoding-cons-rx2} (when both the receivers treat interference as noise) and \ref{sec:rx1-SD-rx2-TIN} (when receiver~$1$ performs successive decoding and receiver~$2$ treats interference as noise) for different values of jamming powers, when the coefficient of self-interference cancelation is $g = 10^{-3}$ or $\beta \triangleq  -10\log g^2 = 60dB$. It can be noticed that, with an increase in the power of the jamming signal, there is also an increase in the throughput for the receiver~$1$. However, the throughput for receiver~$2$ decreases as receiver~$2$ cannot cancel the effect of the jamming power.

\begin{figure}
	\centering
	% \includegraphics[width=3.2in, height=2.1in]{BC_with_queue_secrecy.eps}\\
	\includegraphics[width=3.4in]{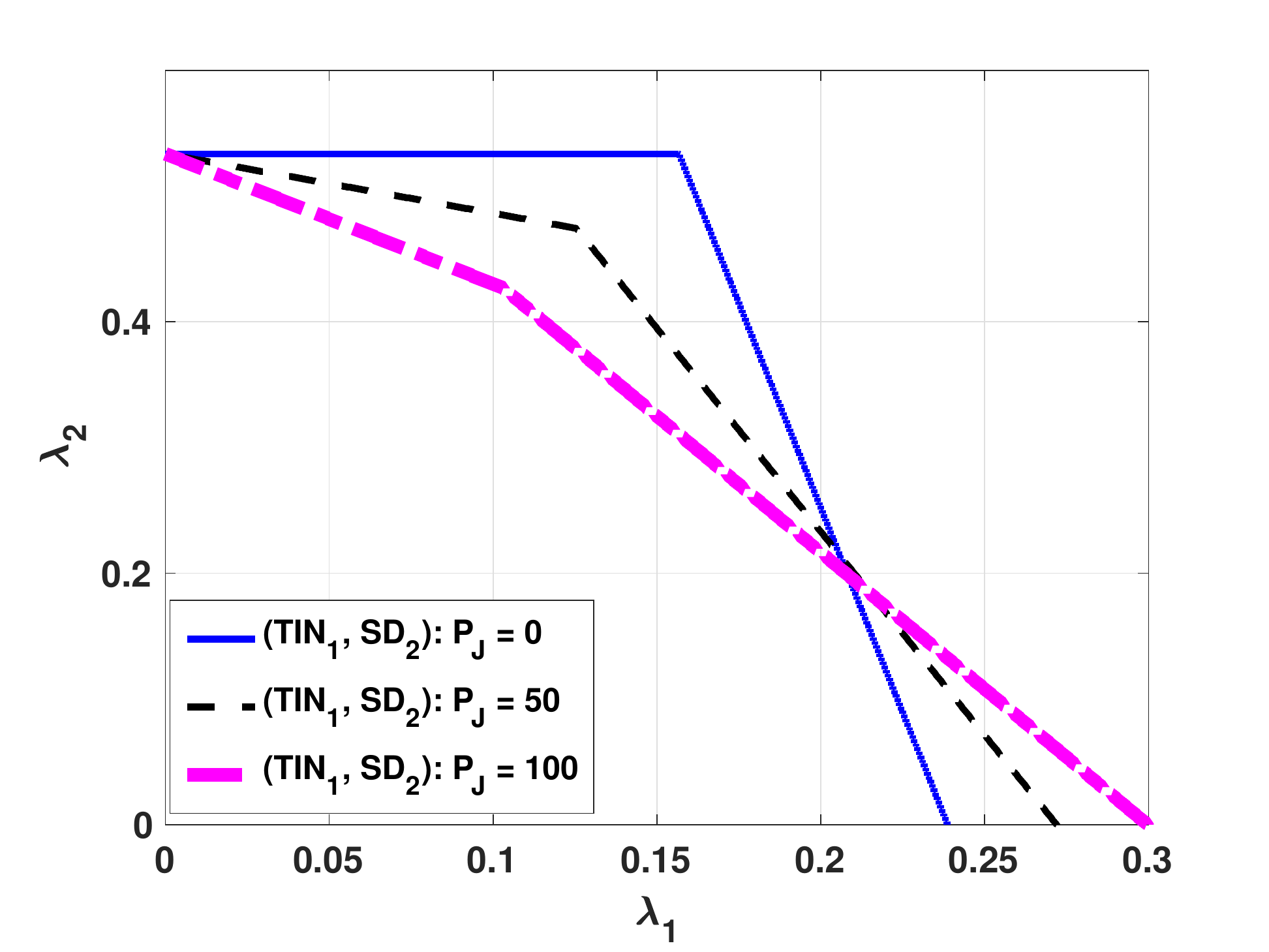}\\
	\caption{The stability region when $Rx-1$ treats interference as noise and Rx~$2$ can perform successive decoding: $P_1 = P_2 = P_J = 100$, $\alpha = 2.8$, $d=10$, $c = 0.8$, $c' = 1.1$, $d_1 = d$, $d_2 = cd$, $d_3 = c'd$, $\gamma_1 = 0.3$, $\gamma_2 = 0.5$, and $\beta = 60dB$. }\label{fig:result5}
\end{figure}

Next we consider the scenario, when receiver~$1$ cannot perform successive decoding. Thus, it treats the packet of the second queue and the residual jamming signal as noise while decoding its intended packet. However, receiver~$2$ can perform successive decoding in this case. The SINR thresholds for decoding packets intended to receiver~$1$ and $2$ are $\gamma_1 = 0.3$ and $\gamma_2= 0.5$, respectively. In Fig.~\ref{fig:result5}, the stability region obtained in Section~\ref{sec:stability-region-rx2} is plotted for $P_1= P_2 = P_J = 100$, and $\beta = 60$dB. From the plot, it can be noticed that it is possible to ensure nonzero stable throughput for queue~$1$, although receiver~$2$ has the capability to perform successive decoding. As the jamming power increases, receiver~$1$ can support larger arrival rates. However, the arrival rate for receiver~$2$ decreases with the increase in the jamming power as it causes additional interference at receiver~$2$.  
\subsection{Impact of Imperfect Self-interference Cancelation on the Stability Region}
\begin{figure}
	\centering
	\includegraphics[width=3.4in]{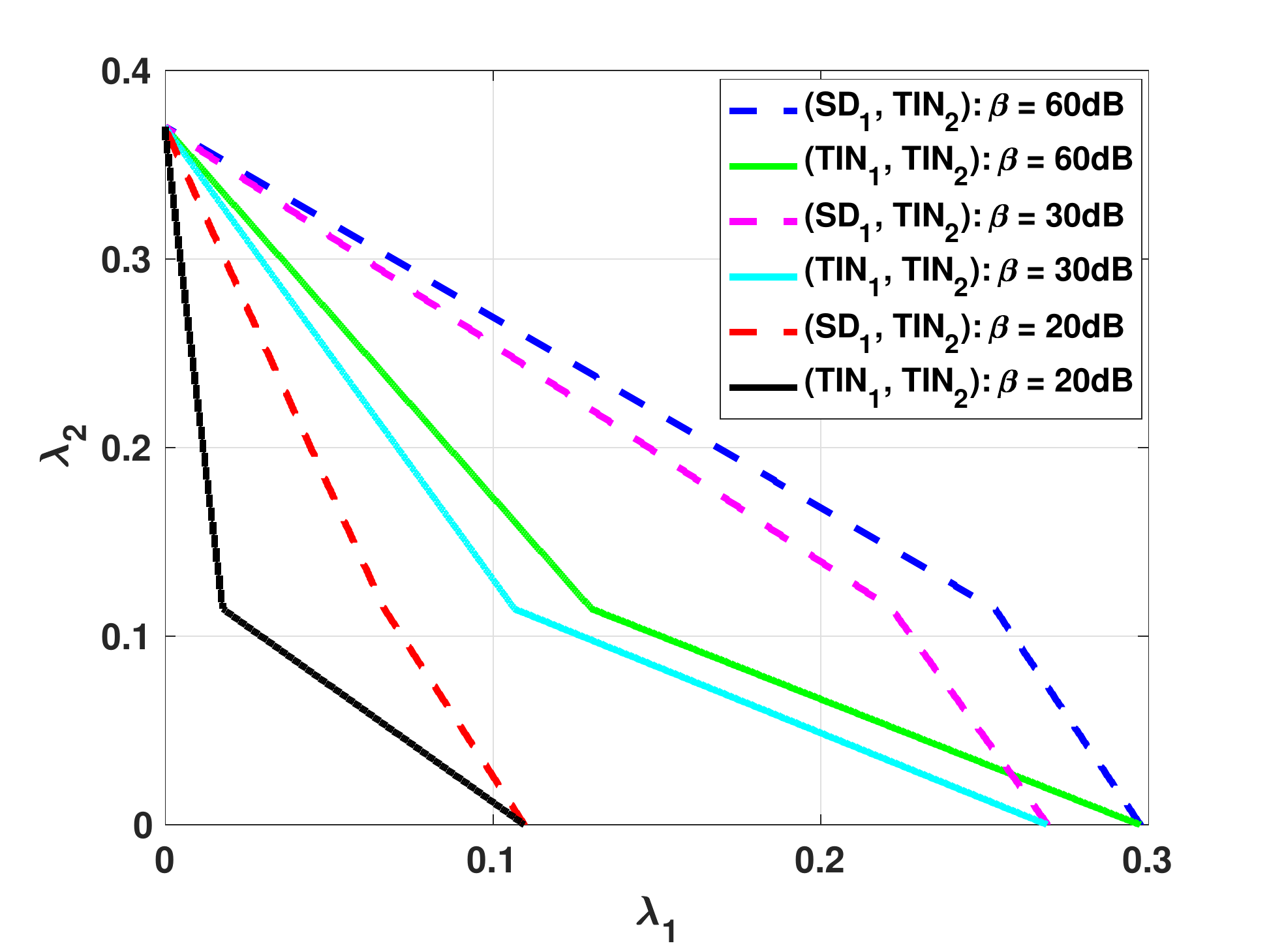} 
	\caption{The stability region for different levels of self-interference isolation: $P_1= P_2= 100$, $P_J= 100$, $\gamma_1=0.5$, $\gamma_2 = 0.4$, $d=10$, $d_1 = d$, $d_2  = d_3 = 1.1d$, and $\alpha= 2.3$.}\label{fig:result4}
	\vspace{-0.4cm}
\end{figure}

In this subsection, the effect of imperfect self-interference cancelation on the stability region is investigated. It is assumed that receiver~$2$ cannot perform successive decoding. In Fig.~\ref{fig:result4}, the stability regions are plotted for different values of coefficient of self-interference cancelation when power of the jamming signal is kept fixed. The stability regions for $(\text{TIN}_1, \text{TIN}_2)$ and $(\text{SD}_1, \text{TIN}_2)$ are obtained using the results in Sections~\ref{sec:decoding-cons-rx2} and \ref{sec:rx1-SD-rx2-TIN}, respectively. It can be noticed that as the value of $\beta$ increases, the stability region decreases. Note that higher the value of $\beta$, better is the self-interference technique at the full-duplex receiver. The shrink in the stability region is more for a small decrease in $\beta$ when the value of $\beta$ is relatively small compared to a higher value of $\beta$. Hence, if the self-interference technique is not efficient, the benefits obtained in service rate due to the ability of receiver~$1$ sending jamming signal to receiver~$2$ can diminish for both the decoding schemes.  

The result for the other case $(\text{TIN}_1, \text{SD}_2)$ can be obtained in a similar manner and  the effect of imperfect self-interference cancelation on the stability region can be more severe due to the fact that receiver~$1$ cannot perform successive decoding as well as there is a secrecy constraint at receiver~$1$.
%\subsection{Closure of the stability region} 
\subsection{Closure of the Stability Region}\label{sec:closure}
In Section~\ref{sec:stability-region-diff-decode}, the stability regions were obtained using success probability for fixed powers at the transmitter and the jammer (receiver~$2$). It was also found that when the self-interference cancelation is not efficient, the gain obtained by jamming can be diminished. A way to overcome this problem is to allocate different transmit powers for the queues at the source and at jammer, which can make the stability region larger. In this subsection, we explore how the stability region can be improved by power allocation. For this, in Figs.~\ref{fig:result6}-\ref{fig:result9}, we plot the closure of stability region, defined as follows. 
\subsubsection{Closure of the stability region over all possible power allocation at the Source} 
Previously, we presented the stability region for given power allocation. In this subsection, we present the stability region over all possible power allocations at the transmitter/source, i.e., $P_1 \in [0, P]$, $P_2 \in [0, P]$, for a given total power constraint, i.e., $P_1 + P_2 = P$. However, the jamming power ($P_J$) is kept fixed. The closure of the stability region over all possible power allocations is defined as
\begin{align} \label{eq:clousure1}
\mathcal{C}(P_J) & \triangleq \left( \bigcup_{ (P_1, P_2) \in [0,P]^2, P_1+P_2=P} \mathcal{C}_1 (P_1,P_2, P_J) \right)  \nonumber \\
& \qquad \bigcup \left( \bigcup_{ (P_1, P_2) \in [0,P]^2, P_1+P_2=P} \mathcal{C}_2 (P_1,P_2, P_J) \right)
\end{align}
where $\mathcal{C}_i (P_1, P_2, P_J) \triangleq \mathcal{R}_i$ for $i=1, 2$. In (\ref{eq:clousure1}), we take the union of the regions over all possible power allocations for each queue corresponding to the closure of the stability region. 

\subsubsection{Closure of the stability region over all possible jamming powers at Receiver 1} 
In this case, the power at the jammer is varied and the power allocation at the transmitter is kept fixed. The closure of the stability region in this case is defined as 
\begin{align} \label{eq:clousure2}
 \mathcal{C}(P_1, P_2) & \triangleq \left( \bigcup_{ P_J \in [0,P_J^{max}]} \mathcal{C}_1 (P_1,P_2, P_J) \right)  \nonumber \\
& \qquad  \bigcup \left( \bigcup_{ P_J \in [0,P_J^{max}]} \mathcal{C}_2 (P_1,P_2, P_J) \right),
\end{align}

\begin{align}
\mathcal{C} & = \lb \displaystyle\cup_{P_J \in [0, P_J^{\max}],  P_1 \text{ and } P_2 \text{ are fixed}}\:\: \mathcal{C}_1 (P_J) \rb \nonumber \\
& \qquad \qquad \cup \lb \displaystyle\cup_{P_J \in [0, P_J^{\max}],   P_1 \text{ and } P_2 \text{ are fixed}}\:\: \mathcal{C}_2 (P_J)\rb, \label{eq:clousure3}
\end{align}
where $\mathcal{C}_i (P_1, P_2, P_J) \triangleq \mathcal{R}_i$ for $i=1, 2$. 

Note that in both cases, $\mathcal{R}_i$ corresponds to the stability region obtained for different decoding schemes in Section~\ref{sec:stability-region-diff-decode}. 

\begin{figure}[t]
	\centering
	% \includegraphics[width=3.2in, height=2.1in]{BC_with_queue_secrecy.eps}\\
	\includegraphics[width=3.4in]{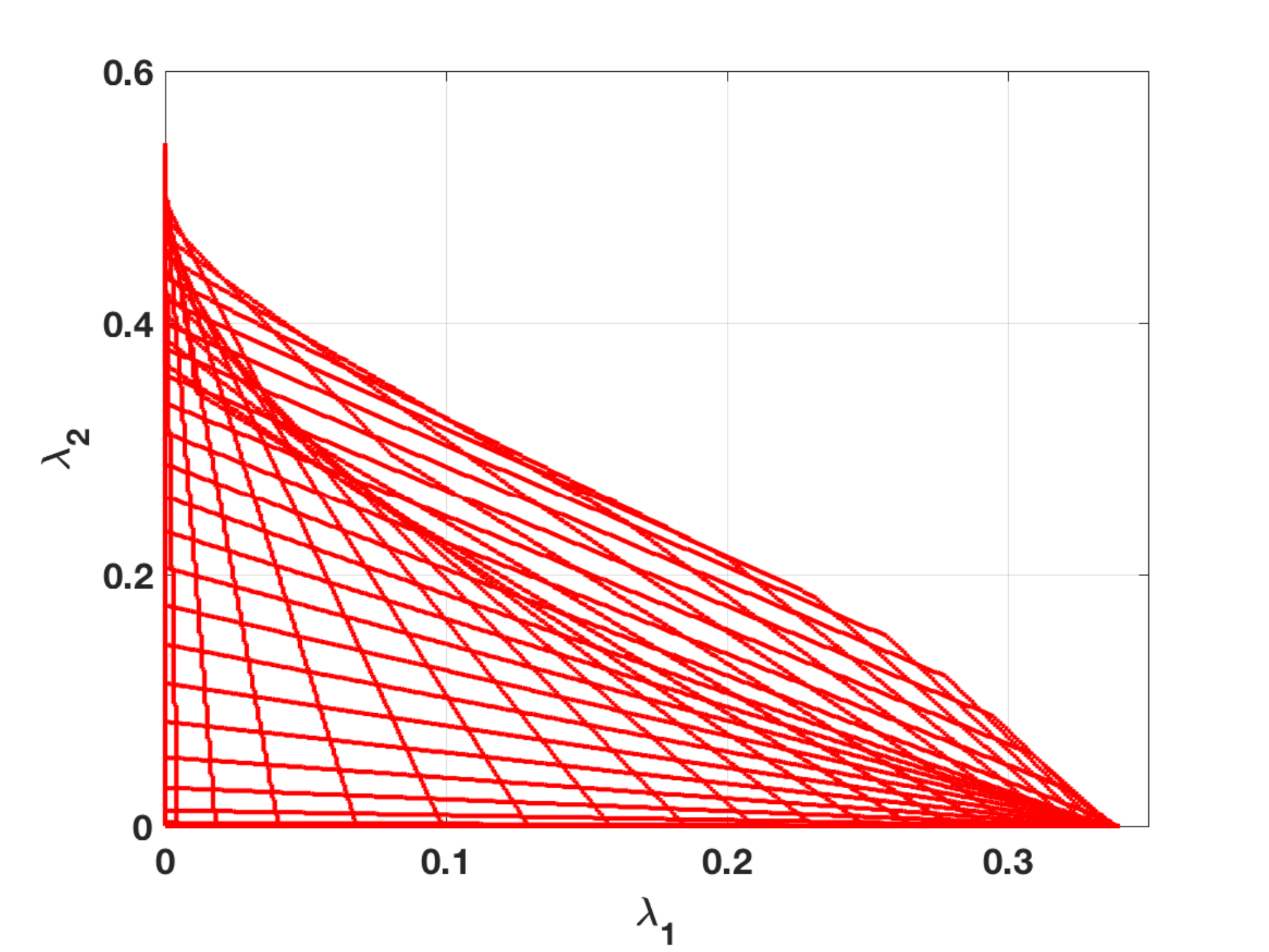}\\
	\caption{Closure of the stability region when both the receivers treat interference as noise: $P = 200$, $P_J = 100$, $\alpha = 2.3$, $d=10$, $c = 1.1$, $c' = 1.2$, $d_1 = d$, $d_2 = cd$, $d_3 = c'd$, $\gamma_1 = 0.4$, $\gamma_2 = 0.35$, and $\beta = 60dB$. }\label{fig:result6}
\end{figure}

\begin{figure}[t!]
	\centering
	% \includegraphics[width=3.2in, height=2.1in]{BC_with_queue_secrecy.eps}\\
	\includegraphics[width=3.4in]{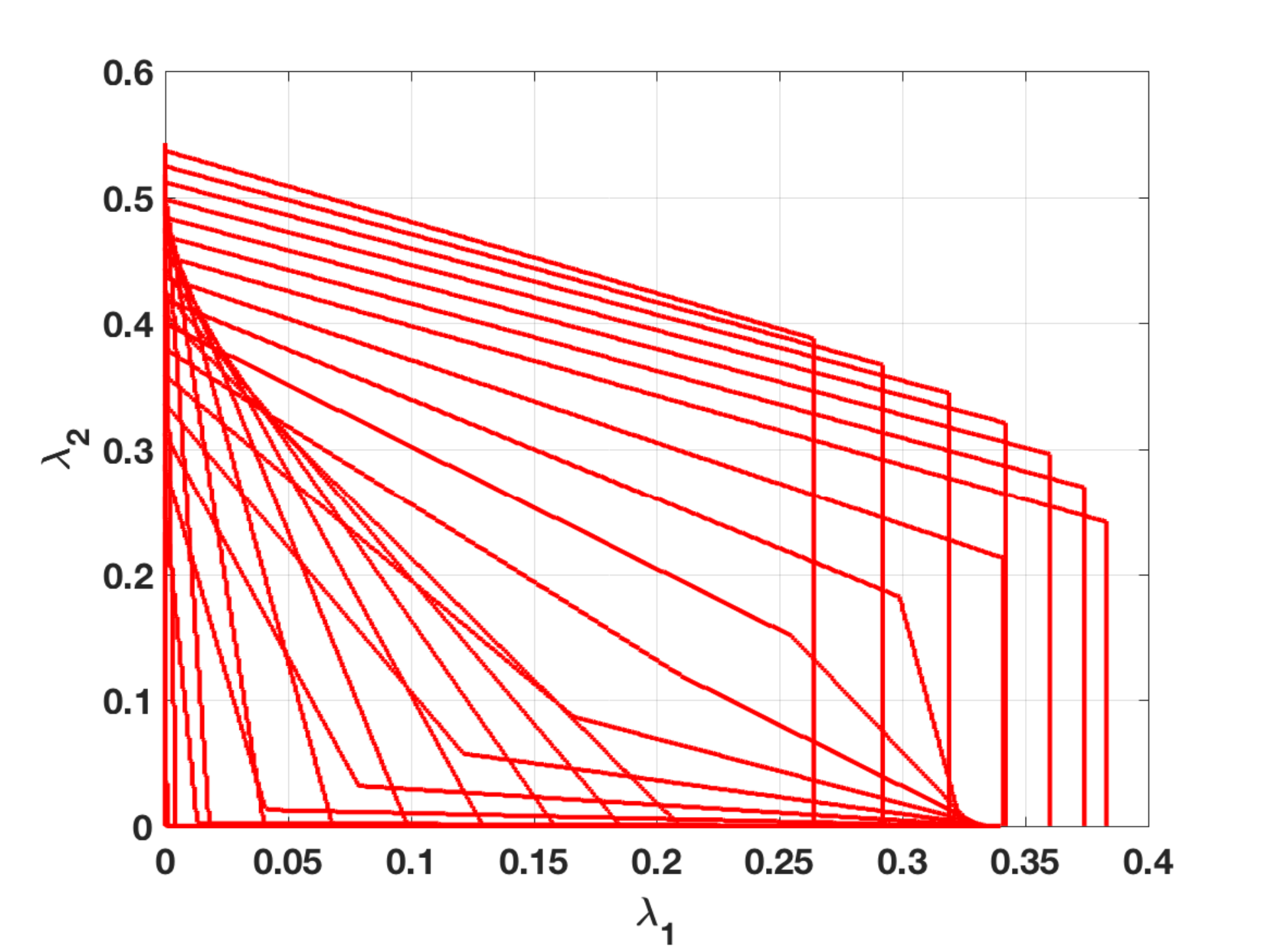}\\
	\caption{Closure of the stability region when receiver~$1$ performs successive decoding while receiver~$2$ treats interference as noise: $P = 200$, $P_J = 100$, $\alpha = 2.3$, $d=10$, $c = 1.1$, $c' = 1.2$, $d_1 = d$, $d_2 = cd$, $d_3 = c'd$, $\gamma_1 = 0.4$, $\gamma_2 = 0.35$, and $\beta = 60dB$. }\label{fig:result7}
\end{figure}

In the following, the closure of the stability region is plotted for different cases assuming that receiver~$2$ cannot perform successive decoding.\footnote{One can also obtain the results for the case when receiver~$2$ performs successive decoding and receiver~$1$ treats interference as noise using the results in Section~\ref{sec:stability-region-rx2}.} First, consider the case, when the transmitter has a total power constraint $P=200$ and the power allocations at both the queues are varied under the following power constraint at the transmitter $P_1 + P_2 = P$. In this case, the jamming power at receiver~$1$ is kept fixed. We obtain the closure of the stability region under two types of decoding schemes: both receivers treat interference as noise (see Fig.~\ref{fig:result6}) and receiver~$1$ performs successive decoding while receiver~$2$ decodes its intended packet by treating other user's packet as noise (see Fig.~\ref{fig:result7}). In both cases, the jamming signal remains as noise at the receivers. From both the plots, it can be noticed that it is possible to achieve a broader stability region with power control at the transmitter. However, power control is not very helpful to enlarge the stability region, when interference is treated as noise at both the receivers. 

\begin{figure}[t!]
	\centering
	% \includegraphics[width=3.2in, height=2.1in]{BC_with_queue_secrecy.eps}\\
	\includegraphics[width=3.4in]{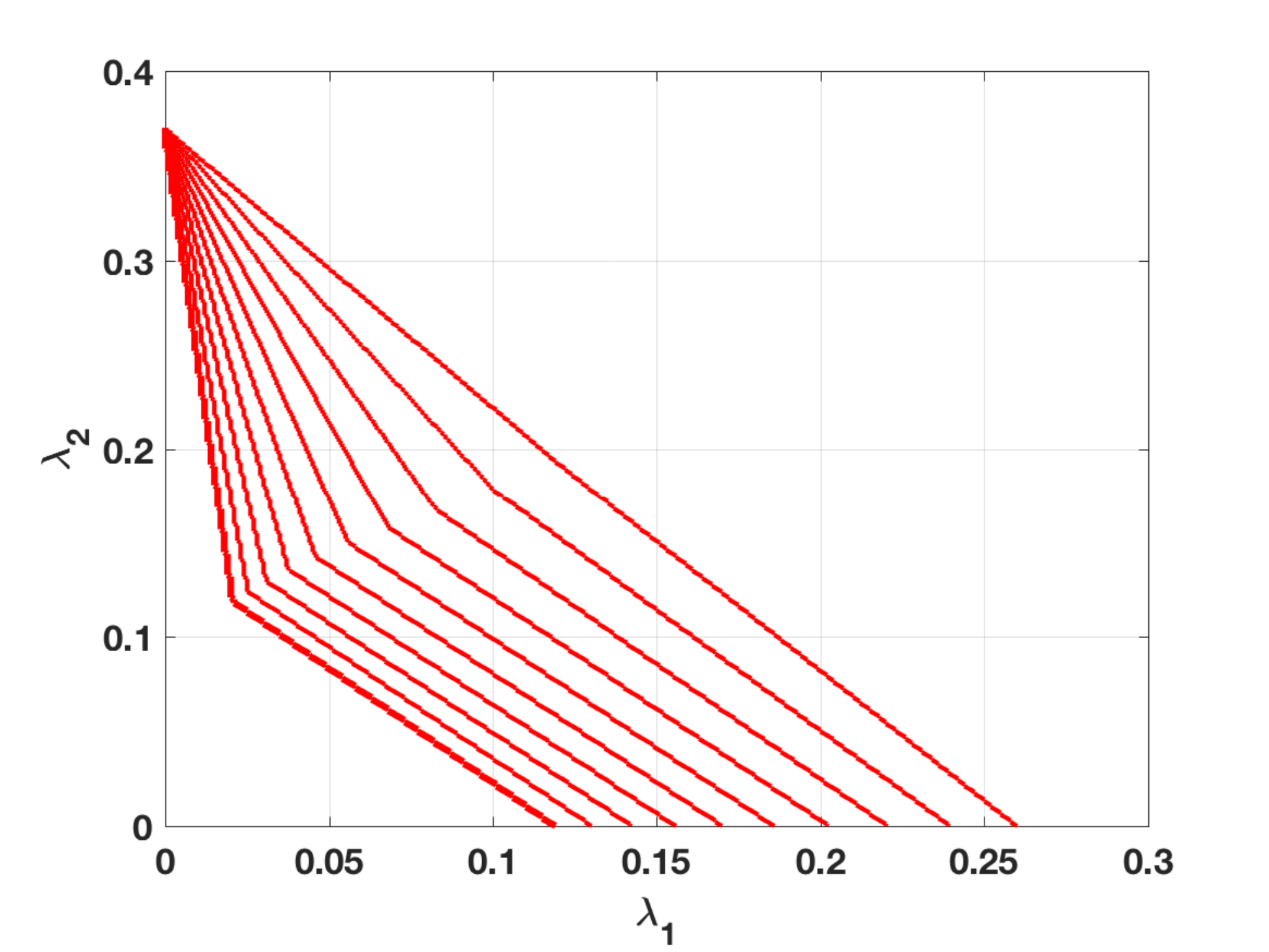}\\
	\caption{Closure of the stability region when both the receivers treat interference as noise: $P_1= P_2 = 100$,  $P_J = 100$, $\alpha = 2.3$, $d=10$, $c = 1.1$,  $d_1 = d$, $d_2 = cd$, $d_3 = cd$, $\gamma_1 = 0.5$, $\gamma_2 = 0.4$, and $\beta = 10dB$.}\label{fig:result8}
\end{figure}
\begin{figure}[t!]
	\centering
	% \includegraphics[width=3.2in, height=2.1in]{BC_with_queue_secrecy.eps}\\
	\includegraphics[width=3.4in]{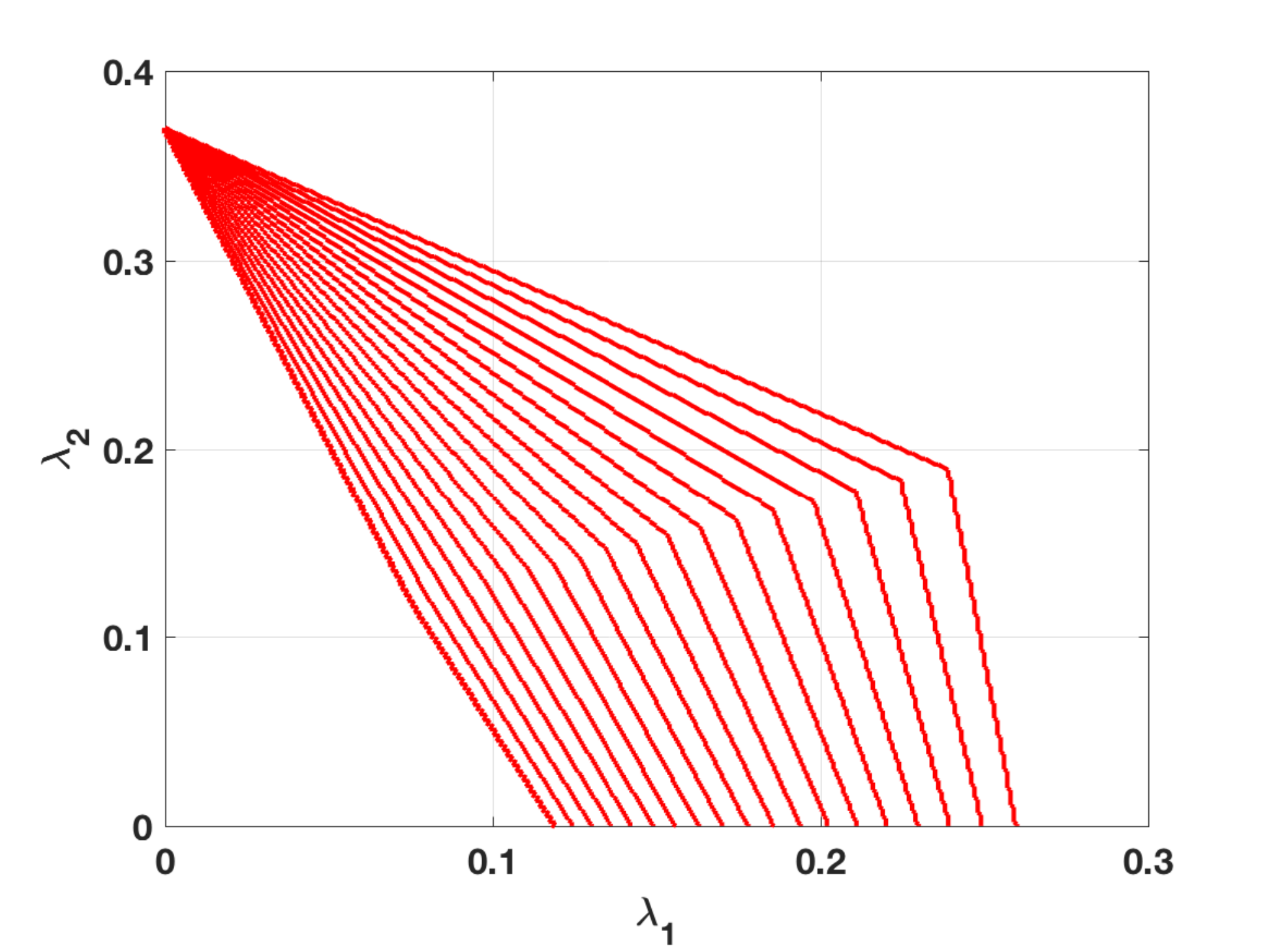}\\
	\caption{Closure of the stability region when receiver~$1$ performs successive decoding while receiver~$2$ treats interference as noise: $P_1= P_2 = 100$,  $P_J = 100$, $\alpha = 2.3$, $d=10$, $c = 1.1$,  $d_1 = d$, $d_2 = cd$, $d_3 = cd$, $\gamma_1 = 0.5$, $\gamma_2 = 0.4$, and $\beta = 10dB$.}\label{fig:result9}
\end{figure}

Now, consider the second case, where the jamming power at receiver~$1$ is varied but fixed power allocation is used at the transmitter for both queues. In this case also, we obtain the closure of the stability region under two types of decoding schemes: both receivers treat interference as noise (see Fig.~\ref{fig:result8}) and receiver~$1$ performs successive decoding while receiver~$2$ decodes its intended packet by treating other user's packet as noise (see Fig.~\ref{fig:result9}). In this case, it can be seen that by performing power control at receiver~$1$, the stability region can be enlarged to a great extent for both schemes compared to Fig.~\ref{fig:result4}. In particular, when the self-interference technique is not very efficient, power control can be quite helpful to enlarge the stability region. 

\subsection{Optimal Jamming Power Design for Maximizing Saturated Throughput of Receiver~$2$} \label{sec:optimization}
In this section, we assume that queue~$2$ always has packets to send while packets arrive at queue~$1$ at a rate $\lambda_1$. As queue~$2$ never empties, we have $\prob(Q_2 > 0) = 1$. The objective here is to maximize the saturated throughput for receiver~$2$ by choosing the jamming power such that a minimum service rate is guaranteed for queue~$1$. The jamming signal transmitted by receiver~$1$ may increase the service rate for queue~$1$ and also can decrease the saturated throughput for receiver~$2$. It is important to note that the self-interference caused at receiver~$1$ may also decrease the service rate seen at the first queue if the self-interference cancelation technique is not efficient. Hence, it is non-trivial to solve the optimization problem. The average packet service rate for the first queue is
\begin{align}
\mu_1 = \prob\lb  \mathcal{D}_{1/1,2}^s \rb.  \label{eq:matsat1}
\end{align}
The average service rate for the second queue is
\begin{align}
\mu_2 = \prob \lb Q_1 > 0\rb \prob\lb  \mathcal{D}_{2/1,2} \rb + \prob \lb Q_1 = 0\rb 
\prob \lb \mathcal{D}_{2/2}\rb, \label{eq:maxsat2}
\end{align}
where $\prob \lb Q_1 > 0 \rb = \frac{\lambda_1}{\prob\lb  \mathcal{D}_{1/1,2}^s \rb}$. Since, the saturated throughput for queue~$2$ is determined by the service rate, it is given by
\begin{align}
\mu_2 = \prob \lb \mathcal{D}_{2/2}\rb - \frac{\prob \lb \mathcal{D}_{2/2}\rb - \prob\lb  \mathcal{D}_{2/1,2}\rb}{\prob\lb  \mathcal{D}_{1/1,2}^s \rb}
\lambda_1. \label{eq:maxsat3}
\end{align}
Finally, the optimization problem can be stated as follows
\begin{align}
& \displaystyle\max_{ P_J }  \quad \prob \lb \mathcal{D}_{2/2}\rb - \frac{\prob \lb \mathcal{D}_{2/2}\rb - \prob\lb  \mathcal{D}_{2/1,2}\rb}{\prob\lb  \mathcal{D}_{1/1,2}^s \rb}\lambda_1, \label{eq:opt1} \\
& \text{s.t.} \quad 0 \leq P_J \leq P_J^{\max}, \text{ and }  \mu_1 \geq \mu_{th}. \label{eq:opt2}
\end{align}
where $P_J^{\max}$ is the maximum power budget at the receiver~$1$. In (\ref{eq:opt2}), the first condition ensures that the power constraint at receiver~$2$ is satisfied and the second condition ensures that the service rate for queue~$1$ is at least $\mu_{th}$. The case of saturated and non-saturated queues can occur in wireless sensor networks where one of the queues receive sensitive information from some specific sensors. Hence, its arrival rate is expected to be random in nature as the sensors may not always have sensitive information to send. The other queue receives general information from many sources and thus, it results in congestion at the second queue. The sensitive information at queue~$1$ needs to be sent to receiver~$1$ under a minimum service rate constraint and without violating the secrecy constraint. The problem considered in this section optimises the jamming power to maximize the throughput of the user without a secrecy constraint under the constraint that the user with the secrecy constraint will be guaranteed to achieve a desired service rate. 

\begin{figure}[t!]
	\centering
	\includegraphics[width=3.4in]{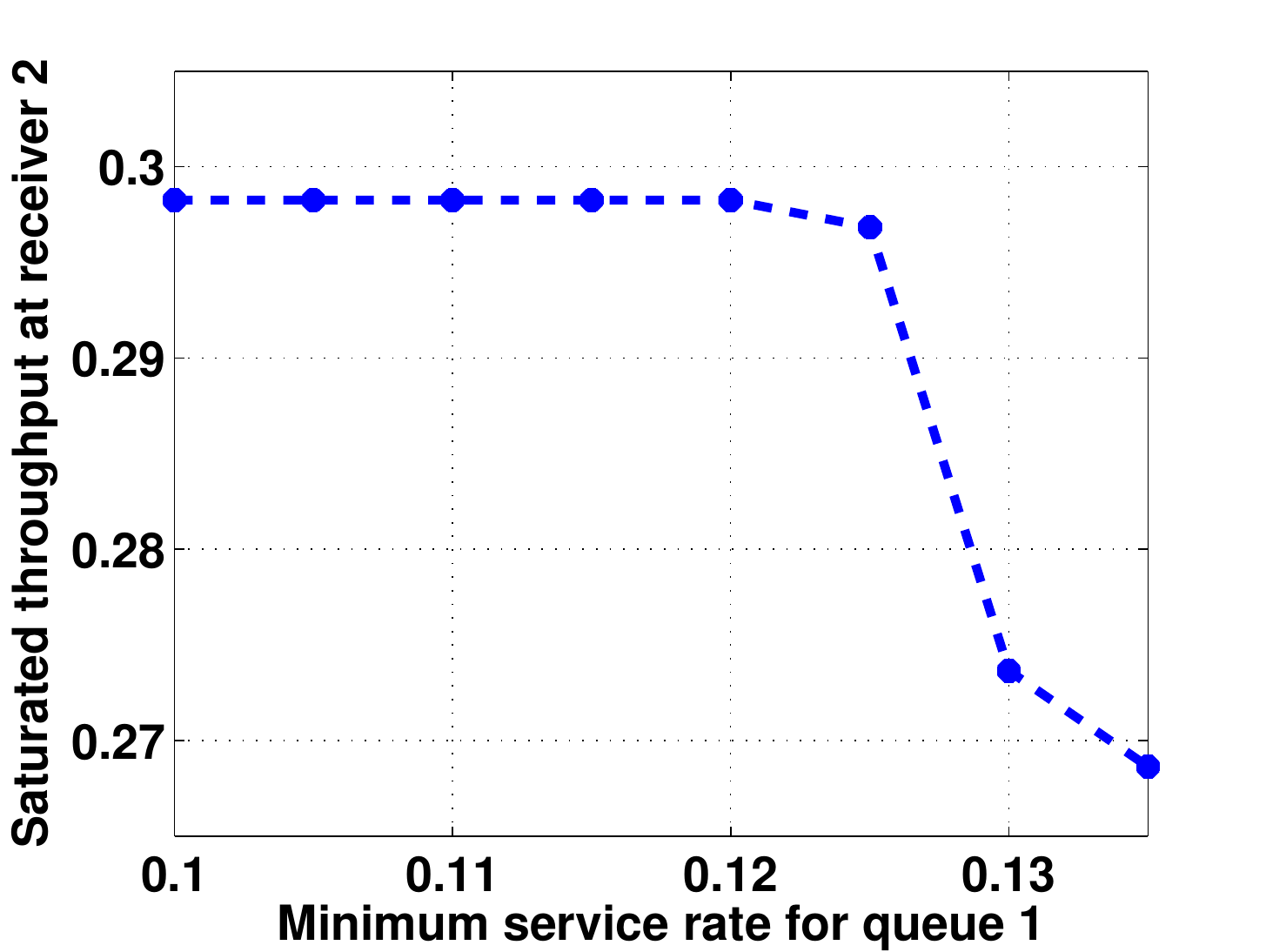}\\
	\caption{Saturated throughput at receiver~$2$ $(\mu_2)$ vs Minimum service rate $(\mu_{th})$: $P_1= P_2 = 100$,  $P_J^{\max} = 100$, $\alpha = 2.2$, $d=10$, $c = 1.1$,  $d_1 = d$, $d_2 = cd$, $d_3 = d$, $\gamma_1 = 0.5$, $\gamma_2 = 0.4$, $\lambda_1 = 0.05$, and $\beta = 80dB$.}\label{fig:result10}
\end{figure}

\begin{figure}[t!]
	\centering
	\includegraphics[width=3.4in]{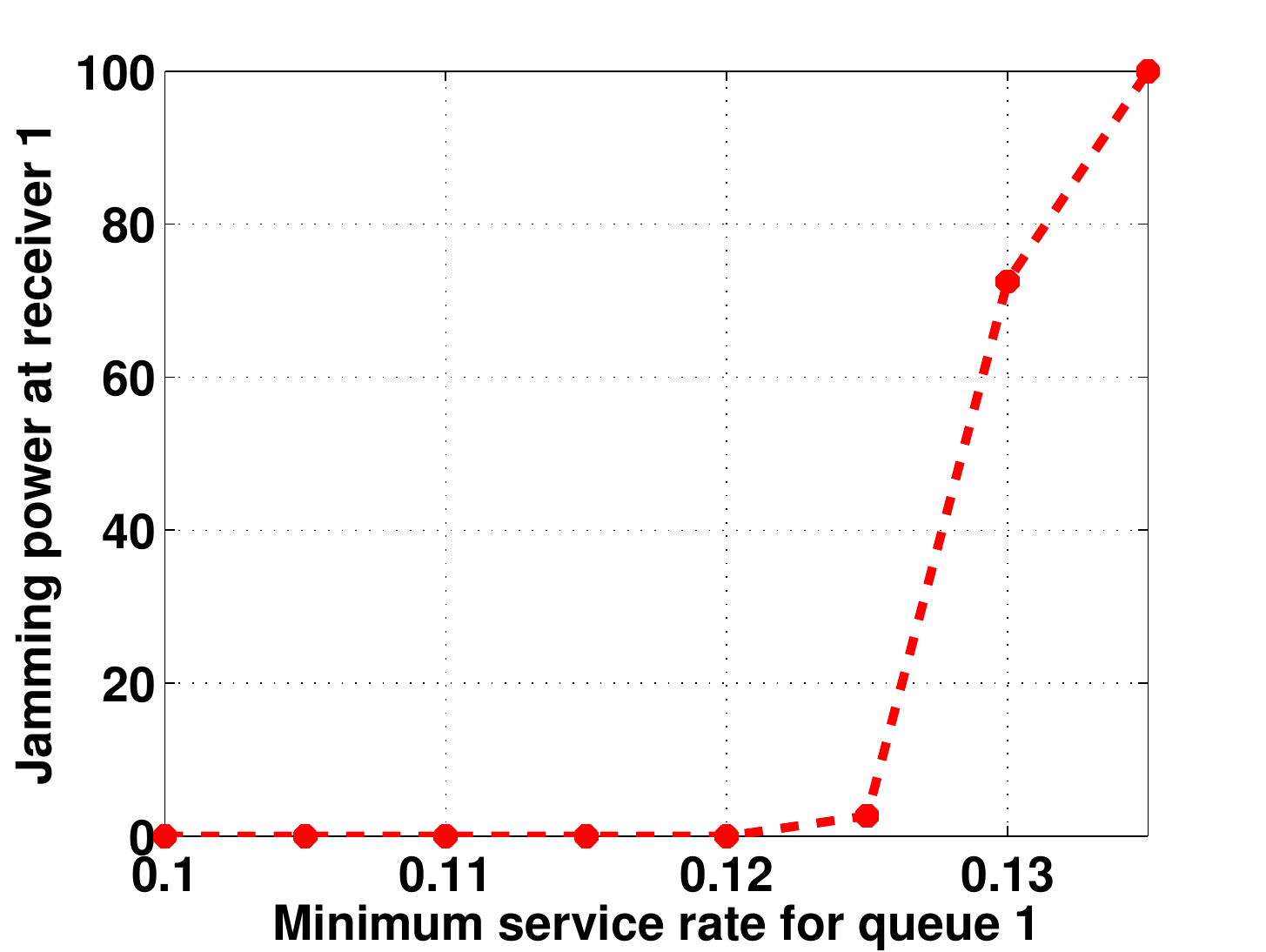}\\
	\caption{Power of the jamming signal sent by receiver~$1$ $(P_J)$ vs Minimum service rate $(\mu_{th})$: $P_1= P_2 = 100$,  $P_J = 100$, $\alpha = 2.3$, $d=10$, $c = 1.1$,  $d_1 = d$, $d_2 = cd$, $d_3 = cd$, $\gamma_1 = 0.5$, $\gamma_2 = 0.4$, $\lambda_1 = 0.05$, and $\beta = 10dB$.}\label{fig:result11}
\end{figure}

In the following, we evaluate numerically the optimization problem in (\ref{eq:opt1}). It is assumed that the individual power allocation at both the queues are $P_1 = P_2 = 100$ and is fixed. The maximum power budget at receiver~$1$ is $P_J^{\max} = 100$. For a desired service rate for queue~$1$, i.e., $\mu_{th}$, we maximize the throughput $\mu_2$ for receiver~$2$ with respect to $P_J$, assuming that queue~$2$ is saturated. In Figs.~\ref{fig:result10} and \ref{fig:result11}, the saturated throughput and the jamming power are plotted versus $\mu_{th}$, respectively. The success probabilities for different cases in (\ref{eq:opt1}) are evaluated using the results derived in Sec.~\ref{sec:decoding-cons-rx2}. From the plots, it can be noticed that when $\mu_{th} \leq 0.125$, receiver~$1$ does not need to send a jamming signal to guarantee minimum service rate for queue~$1$. However, when $\mu_{th} > 0.125$, receiver~$1$ needs to send a jamming signal of different power levels to guarantee minimum service rate for queue~$1$ as shown in Fig.~\ref{fig:result11}. This leads to decrease in the saturated throughput for receiver~$2$ as $\mu_{th}$ increases beyond $0.125$. 

\section{Conclusion}
In this work, the stability region of the two-user broadcast channel is studied, where one of the receivers with secrecy constraint has full-duplex capability. The stability region of the BC is obtained for the general case. Then, the stability region is obtained for different decoding schemes at the receivers. The analysis takes into account the secrecy constraint at the receiver and impact of the imperfect self-interference cancelation on the stability region simultaneously. When receiver~$2$ can perform successive decoding, the role of jamming signal to ensure secrecy of the packets intended to receiver~$1$ is more crucial. It is also found that performing power control at the transmitter or receiver~$1$ can enlarge the stability region significantly. When the self-interference cancelation is not efficient, performing power control at receiver~$1$ can help to increase the stability throughput significantly. From the optimization problem considered to maximize the saturated throughput of second queue while guaranteeing minimum service rate for the first queue, it is also found that the jamming capability of receiver~$1$ can help to guarantee minimum service rate and at the same time can maximize the saturated throughput of receiver~$2$ by appropriately choosing the jamming power. 
  
\bibliographystyle{IEEEtran}
\bibliography{IEEEabrv,bib_stability_region_ver0}
\end{document}